\documentclass[journal=jctcce,manuscript=article]{achemso}
\usepackage{amssymb,amsmath,amsfonts,multicol,multirow,longtable,array,mathpazo}
\usepackage{lscape}
\usepackage[version=3]{mhchem} 
\usepackage{appendix}
\usepackage{soul} 
\usepackage[usenames]{color}
\usepackage{makecell}
\usepackage{enumerate}
\usepackage{xr}
\usepackage[T1]{fontenc} 
\usepackage{booktabs}
\usepackage{tikz}
\usetikzlibrary{shapes.geometric}
\usepackage{threeparttable}
\usepackage{graphicx}
\usepackage{subfigure}
\usepackage{colortbl}
\usepackage{framed}
\usepackage{verbatim}
\usepackage{amsbsy}
\usepackage{bm}
\usepackage{bbm}
\usepackage[normalem]{ulem}
\usepackage{float}
\usepackage[linesnumbered,boxed]{algorithm2e}
\usepackage{algorithm2e}
\usepackage{float}
\usepackage{subfigure}
\usepackage{simplewick} 

\newcounter{xscheme}

\newfloat{Algorithm}{htbp}{alg}
\floatname{Algorithm}{Algorithm}

\newcounter{exe}[figure]
\newcommand{\iexe}{\refstepcounter{exe}\the\value{exe}:}

\setkeys{acs}{maxauthors = 0} 

\author{Chunzhang Liu}
\author{Ning Zhang}
\author{Wenjian Liu}\email{liuwj@sdu.edu.cn}
\affiliation{Qingdao Institute for Theoretical and Computational Sciences and Center for Optics Research and Engineering,
	Shandong University, Qingdao, Shandong 266237, China}

\title{PASPT2: A Size-Extensive and Size-Consistent Partial-Active-Space Multistate Multireference Second-Order Perturbation Theory
for Strongly Correlated Electrons}

\setkeys{acs}{articletitle=true}

\begin{document}
	
\begin{abstract}
A partial-active-space (PAS) multi-state (MS) multi-reference second-order perturbation theory (MRPT2) for
the electronic structure of strongly correlated systems of electrons,
dubbed PASPT2, is formulated by linearizing the intermediate normalization-based general-model-space state-universal
coupled-cluster theory with singles and doubles [IN-GMS-SU-CCSD; J. Chem. Phys. 119, 5320 (2003)].
At variance with the existence of disconnected terms in the IN-GMS-SU-CCSD amplitude equations,
the disconnected terms in the PASPT2 amplitude equations can be
avoided completely by choosing a special reference-specific zeroth-order Hamiltonian.
The corresponding effective/intermediate Hamiltonian can also be made connected and closed, so as to render the energies
obtained by diagonalization fully connected. As such, PASPT2 is strictly size-extensive,
in sharp contrast with the parent IN-GMS-SU-CCSD.
It is also size-consistent when the PAS of a supermolecule
is chosen to be the direct product of those of the physically separated, non-interacting fragments.
Prototypical systems are taken as showcases to reveal the efficacy of PASPT2.

\end{abstract}


\noindent
Precision Chemistry, DOI: 10.1021/prechem.5c00408 (2026).

\maketitle

\clearpage
\newpage

\section{INTRODUCTION}
Accurate descriptions of the electronic structures of strongly correlated systems of electrons remain a great
challenge to quantum chemistry. Such systems (e.g., polynuclear transition metal complexes, etc.) are characteristic of many valence electrons and very
many valence orbitals that are energetically near-degenerate and spatially proximal. As a direct result, the wave function appears as a
massive superposition of electron configurations (Slater determinants or configuration state functions), none of which plays the leading role.
Such systems also feature many low-lying states of different spins, so that even a qualitative description already
requires the use of many configurations.

Following the common nomenclature, the component of electron correlation
furnished by the most important configurations is called static (or nondynamic) correlation, whereas that due to the remaining configurations is dynamic in nature.
Such a partitioning of electron correlation is of course arbitrary. However, the concept is very useful for
interpretation and even for design of new wave function Ans\"atze. The very first question is then how to
pick up those ``most important configurations'' for static correlation. Chemical and physical intuitions are certainly very helpful here. However,
the manually chosen configurations may not all be important for all molecular geometries. One way out is to pick up all configurations that
can be generated by distributing $N$ valence electrons in a prechosen set of $N_a$ valence (or active) orbitals in all possible ways.
Such an approach is called complete active space configuration interaction (CASCI) or CASSCF if the orbitals
are further optimized self-consistently (SCF)\cite{RuedenbergCAS,MCSCF0,MCSCF1,MCSCF2,MCSCF3}.
The latter has been a cornerstone of quantum chemistry ever since its introduction.
However, CASCI/CASSCF is not free of problems. Among others, it scales combinatorially  with respect to the number of active orbitals, so as
to limit its applicability to just a dozen of active orbitals. More severely, a CAS may contain many configurations that
have little contribution to the wave function. Such configurations may even be higher in energy than some of the external space.
Treating the former exactly (by diagonalization) whereas the latter approximately (by perturbation)
is obviously unbalanced. In particular, the former are the source of intruder states\cite{Intruder1972} that plague
non-variational methods, especially in the treatment of high-lying electronic states\cite{CASPT2Intruder,CASPT2IntruderImaginary,NEVPT2Intruder1,NEVPT2Intruder2}.
A more appealing approach is to perform an iterative selection of configurations, so as to
end up with a compact variational wave function that is adaptive to the nature of the system under consideration.
The residual (inner-space) dynamic correlation can then be well described by a second-order perturbation theory (PT2).
Such approaches can be termed collectively as sCIPT2 (selected CI plus PT2), which have a long history\cite{Davidson1969,Whitten1969,PeyerimhoffsCI1974,CIPSIa,CIPSIb}
and have been revived in the last decade\cite{LambdaCI2014,SDS,iCI,ACI2016,ACI2017,HBCI2016,HBCI2017a,ASCI2016,ASCI2018PT2,iFCI2017a,iFCI2017b,iCIPT2,iCIPT2New,iCISCF,ICEa2021,ICEb2021}.
In particular, when the selection of configurations is performed over the whole Hilbert space, sCIPT2 combined with extrapolations
can provide near-exact results\cite{benzene}. Nevertheless, if the selection is not performed over the whole Hilbert space but over, e.g., a large CAS,
some size-extensive multi-reference second-order perturbation theory (MRPT2) remains to be developed on top of sCI,
for the Epstein-Nesbet type\cite{Epstein,Nesbet} of PT2
employed in sCIPT2 is not suited for the dynamic correlation outside such selected space.
To put it in more general terms, a size-extensive state-specific (SS) or multi-state (MS)
MRPT2 based on a general model space (GMS) needs to be formulated, which
is distinct from those counterparts that are based on
a structured reference (e.g., CAS, antisymmetrized product of strongly orthogonal geminals or perfect pairing generalized valence bond)
\cite{SS-MRPT2a,SS-MRPT2b,MaoSS-MRPT2a,MaoSS-MRPT2b,UGA-SSMRPT2,DSRG-MRPT22015,SA-DSRG-MRPT22018,JM-MRPT2,NEVPT2a,NEVPT2c,ChenFW2014,ChenFW2019,
APSG-H2PT2,UNOSLGPT22019,GVB-BCPT22013}.

At this moment, some terminologies spread in the literature need to be clarified. Instead of the widely used CAS, the term `complete model space' (CMS)
is also in use. The two are obviously identical.
Here, the whole set of orbitals is partitioned into three disjoint subsets: core, active and external orbitals that are
always occupied, variably occupied and always unoccupied, respectively, in all configurations of CAS/CMS.
The counterpart of CAS/CMS is incomplete active space (IAS), incomplete model space (IMS) or partial active space (PAS) that is only a subset of a CAS/CMS.
In contrast, the term GMS does not rely on
any partitioning of the orbitals. Rather, it can be composed of arbitrary electron configurations. However, it is not always
possible to choose manually the relevant configurations. Therefore, GMS is in practice the same as IAS/IMS/PAS
that can readily be constructed by performing a configuration selection over a CAS. To differentiate from the size-inextensive GMSPT2 approach\cite{GMSPT2a,GMSPT2b,GMSPT2c},
the particular variant of GMS/IAS/IMS/PAS-based MS-MRPT2 proposed here is to be dubbed PASPT2, which is
not only size-extensive but also size-consistent when the PAS of a supermolecule
is the direct product of those of the physically separated, non-interacting fragments, expressed in terms of orbitals localized on the fragments.

The remaining of the paper is organized as follows. After a brief recapitulation of the
intermediate normalization-based general-model-space state-universal coupled-cluster theory (IN-GMS-SU-CC)\cite{GMS-SU-CC2003}
in Sec. \ref{SecSUCC}, PASPT2 is formulated in Sec. \ref{SecPASPT2} by linearizing IN-GMS-SU-CCSD (coupled-cluster with singles and doubles),
where the size-extensivity and size-consistency of PASPT2 are addressed in detail.
The implementation and pilot applications are then provided in Secs. \ref{Implementation} and
\ref{Pilot}, respectively. The paper is finally closed with concluding remarks in Sec. \ref{Conclusion}.

\section{THEORY}\label{SecTheory}
\subsection{IN-GMS-SU-CC}\label{SecSUCC}
For the purpose of introducing the notations, we first recapitulate
the IN-GMS-SU-CC theory of Li and Paldus\cite{GMS-SU-CC2003}, which is based on the Jeziorski-Monkhorst
Ansatz\cite{JM1981} for MRCC. Briefly, given a PAS $\mathcal{M}_0$ spanned by $M$
$N$-electron determinants (NED) $\{|\alpha\rangle, |\beta\rangle, |\gamma\rangle, \cdots \}$, Jeziorski and Monkhorst\cite{JM1981}
introduced a multi-exponential form for the wave operator
\begin{align}
\Omega&=\sum_{\alpha\in\mathcal{M}_0}e^{T^\alpha}P_{\alpha}=\Omega P,\label{Omega}\\
P&=\sum_{\alpha\in\mathcal{M}_0}P_{\alpha},\quad P_{\alpha}=|\alpha\rangle\langle\alpha|,
\end{align}
to map the PAS $\mathcal{M}_0$ onto the exact manifold $\mathcal{M}$ spanned by $M$
exact solutions of the (second-quantized) Schr\"odinger equation (SEQ).
Since the action of $\Omega$ on the
complementary space $\mathcal{M}_0^\perp$ characterized by the projector $Q=1-P$ is of no interest at all, $\Omega Q$ can simply be set to zero, as already implied by
the second equality of Eq. \eqref{Omega}. The action of $\Omega$ on the reference NED $|\alpha\rangle$
reads
\begin{align}
\Phi_{\alpha}=\Omega |\alpha\rangle=P\Omega |\alpha\rangle + Q \Omega |\alpha\rangle,
\end{align}
where the first term is merely a remixing of the reference NEDs [i.e., $\sum_{\beta\in \mathcal{M}_0}|\beta\rangle \langle \beta|\Omega|\alpha\rangle$]
and hence does not take an effective step toward the target solutions. Because of this, one is free to impose the so-called
intermediate normalization condition (INC)
\begin{align}
P\Omega P&=P,\label{POP}\\
\langle \beta|\Omega|\alpha\rangle&=\langle \beta|e^{T^\alpha}|\alpha\rangle=\delta_{\beta\alpha},\label{INC}
\end{align}
so as to obtain the following perturbed functions
\begin{align}
\Phi_{\alpha}=\Omega |\alpha\rangle=|\alpha\rangle + Q \Omega |\alpha\rangle,
\end{align}
which can be taken as the basis to expand
the exact eigenfunctions $\{\Psi_\alpha\}_{\alpha=1}^M$, i.e.,
\begin{align}
\Psi_\alpha&=\sum_{\beta\in \mathcal{M}_0} \Phi_{\beta}C_{\beta \alpha}=\Omega \tilde{\Phi}_\alpha,\label{ExactWF}\\
\tilde{\Phi}_\alpha&=\sum_{\beta\in \mathcal{M}_0} |\beta\rangle C_{\beta \alpha}=P\Psi_\alpha.\label{Bonnes}
\end{align}
It follows that, under the INC \eqref{POP}, $P$ $(=P^2)$ and $\Omega$ $(=\Omega^2)$ are the
one-to-one mapping from $\mathcal{M}$ to $\mathcal{M}_0$ and the inverse one-to-one mapping
from $\mathcal{M}_0$ to $\mathcal{M}$, respectively. Of course, such mappings hold true only if the \emph{bonnes fonctions}\cite{Bloch1958}
$\{\tilde{\Phi}_\alpha\}_{\alpha=1}^M$ (and hence
$\{\Psi_\alpha\}_{\alpha=1}^M$) are linearly independent. This is equivalent to the requirement
that none of the exact functions $\{\Psi_\alpha\}_{\alpha=1}^M$ is orthogonal to all the reference NEDs\cite{JM1981}.

Plugging the expansion \eqref{ExactWF} into the SEQ
\begin{align}
H\Psi_\alpha&=E_\alpha \Psi_\alpha,
\end{align}
namely,
\begin{align}
H\Omega \tilde{\Phi}_\alpha&=E_\alpha \Omega \tilde{\Phi}_\alpha,\label{SEQ}
\end{align}
and projecting on the left with $P$ give rise to the effective, nonhermitian SEQ
\begin{align}
H^{\mathrm{eff}} \tilde{\Phi}_\alpha&=E_\alpha \tilde{\Phi}_\alpha,\quad H^{\mathrm{eff}} =PH\Omega P,\label{Heff}
\end{align}
or in matrix form
\begin{align}
\mathbf{H}^{\mathrm{eff}} \mathbf{C}&=\mathbf{C} \mathbf{E}, \quad
 H^{\mathrm{eff}}_{\beta\alpha}=\langle\beta|H\Omega|\alpha\rangle.\label{HeffMat}
\end{align}
Eq. \eqref{SEQ} can then be rewritten as
\begin{align}
H \Omega \tilde{\Phi}_\alpha = E_\alpha \Omega \tilde{\Phi}_\alpha =\Omega H^{\mathrm{eff}} \tilde{\Phi}_\alpha.
\end{align}
This equation must hold for all $\{\tilde{\Phi}_\alpha\}$ as well as their linear combinations, thereby leading to
the generalized Bloch equation\cite{LindgrenBook}
\begin{align}
 H \Omega= \Omega H^{\mathrm{eff}},\label{GenBloch}
\end{align}
which reads more explicitly in the present context
\begin{align}
\sum_{\beta\in \mathcal{M}_0} H e^{T^\beta} P_\beta = \sum_{\beta\in \mathcal{M}_0} e^{T^\beta}P_\beta H^{\mathrm{eff}} P.
\end{align}
Projecting on the left with $e^{-T^\alpha}$ and on the right with $P_\alpha$, we obtain an alternative form of Eq. \eqref{GenBloch}
\begin{align}
e^{-T^\alpha} H e^{T^\alpha} P_\alpha
= \sum_{\beta\in \mathcal{M}_0} e^{-T^\alpha} e^{T^\beta}P_\beta H^{\mathrm{eff}} P_\alpha.\label{GenBlochT}
\end{align}

At this stage, the $T^\alpha$-cluster operator remains to be defined.
A salient feature of the wave operator $\Omega$ defined in Eq. \eqref{Omega} lies in that it takes each reference NED as a distinct Fermi vacuum, so that
the parametrization of the $T^\alpha$-cluster operator can proceed in the same way as in single-reference (SR) CC, i.e.,
\begin{align}
T^\alpha&=T^\alpha_1+T^\alpha_2+\cdots\label{T-clusters}\\
&=t^i_a(\alpha)\{a^a_i\}_\alpha+\frac{1}{4} t^{ij}_{ab}(\alpha)\{a^{ab}_{ij}\}_\alpha+\cdots\\
a^a_i&=a_a^\dag a_i,\quad a^{ab}_{ij}=a_a^\dag a_b^\dag a_j a_i,
\end{align}
where (and throughout the text) the Einstein summation convention over repeated indices is employed. The
$\mathbf{t}$-amplitudes are antisymmetric and nonzero only when the spin orbitals $\{i,j,\cdots\}$
and $\{a,b,\cdots\}$ are occupied and unoccupied, respectively, in the reference NED $|\alpha\rangle$.
However, the INC \eqref{INC} dictates that the internal mapping (i.e., the action of $\Omega$ on a reference NED gives another
reference NED) must be avoided. In case that $\mathcal{M}_0$ is complete (i.e.,
invariant under unitary transformations of the active orbitals),
this is equivalent to setting all internal $T^\alpha$-cluster operators (which excite one reference NED to another) to zero.
This is because a CASCI wave function $\Psi_I$ can always be recast into an exponential form by taking any NED (e.g., $|\alpha\rangle$) of $\mathcal{M}_0$
as the reference, i.e.,
\begin{align}
 |\Psi_I\rangle&=e^{\sigma}|\alpha\rangle =\sum_{\beta\in\mathrm{CAS}} |\beta\rangle \langle\beta |e^{\sigma^\alpha_1+\cdots+\sigma^\alpha_N}|\alpha\rangle\nonumber\\
 &= \sum_{\beta\in\mathrm{CAS}} |\beta\rangle \langle\beta | 1+\tilde{C}_1^\alpha +\cdots+\tilde{C}_N^\alpha|\alpha\rangle,\label{CASCIWF}
\end{align}
 where
\begin{align}
 \tilde{C}_1^\alpha&=\sigma_1^\alpha,\\
 \tilde{C}_2^\alpha&=\sigma_2^\alpha + \frac{1}{2}(\sigma_1^\alpha)^2,\\
 \tilde{C}_3^\alpha&=\sigma_3^\alpha + \sigma_1^\alpha\sigma_2^\alpha+\frac{1}{3!}(\sigma_1^\alpha)^3, \mbox{ etc.}
\end{align}
It follows that any term in the internal connected cluster $\sigma^\alpha_n$ that generates a reference NED $|\beta\rangle$ is accompanied by
disconnected-cluster terms generating the same $|\beta\rangle$, expressed as products of lower-order \emph{internal}
excitations. However, the situation is different when
the model space $\mathcal{M}_0$ is incomplete: some lower excitations in the disconnected-cluster terms
may generate \emph{external} ($Q$-space) functions, so that the corresponding connected-cluster amplitudes cannot simply be set to zero
without introducing approximations.
Based on this observation, Li and Paldus\cite{GMS-SU-CC2003}
introduced the so-called connectivity condition (or C-condition), which
set the amplitudes $\mathbf{C}(\alpha)$ of the internal $C_n^\alpha$ operators in $e^{T^\alpha}$ ($=1+C_1^\alpha+\cdots+C_N^\alpha$),
instead of the amplitudes of the internal $T^\alpha$-clusters, to zero.
For instance, if the operator $a_i^a$/$a_{ij}^{ab}$ generates a reference NED when operating on $|\alpha\rangle$,
we will have
\begin{subequations}\label{C-Cond}
\begin{equation}
C^i_a(\alpha)=t^i_a(\alpha)=0,\label{1BCond}
\end{equation}
\begin{equation}
C^{ij}_{ab}(\alpha)=t^{ij}_{ab}(\alpha)+t^{i}_{a}(\alpha) t^{j}_{b}(\alpha)-t^{i}_{b}(\alpha) t^{j}_{a}(\alpha)=0,\label{2BCond}
\end{equation}
\end{subequations}
where $t^{ij}_{ab}(\alpha)$ would be nonzero if $a_{i}^{a}$ and $a_{j}^{b}$ and/or $a_{i}^{b}$ and $a_{j}^{a}$
are external excitations for $|\alpha\rangle$. Having determined the amplitudes of the internal $T^\alpha$-cluster operators this way, those of the external $T^\alpha$-cluster operators
can be obtained by applying an external NED $\langle \chi_{l\alpha}|=(a^{a,b,\cdots}_{i,j,\cdots}|\alpha\rangle)^\dag$ on the left and the reference NED $|\alpha\rangle$ on the right of
Eq. \eqref{GenBlochT},
 \begin{align}
\langle\chi_{l\alpha}| e^{-T^\alpha} H e^{T^\alpha} |\alpha\rangle
&= \sum_{\beta \ne\alpha\in \mathcal{M}_0} C_{l\alpha,\beta}
\langle\beta| H^{\mathrm{eff}} |\alpha\rangle,\label{SU-CC-T}\\
C_{l\alpha,\beta}&=\langle\chi_{l\alpha}| e^{-T^\alpha} e^{T^\beta}|\beta\rangle, \label{Coupling}
 \end{align}
The left-hand side (LHS) of Eq. \eqref{SU-CC-T} is of the same form as in SR-CC, whereas the right-hand side (RHS) is characteristic of the multi-reference case.

Eq. \eqref{SU-CC-T} is independent of the amplitudes of the target states $\{\Psi_\alpha\}$ and hence state-universal. It means that
this system of nonlinear equations has to be solved for the $M$ target states simultaneously,
which is a tremendous task. Taking IN-GMS-SU-CCSD as an example,
the LHS of Eq. \eqref{SU-CC-T} scales as $\mathcal{O}(M K_o^2 K_v^4) \sim \mathcal{O}(M K^6)$,
whereas both the coupling and effective Hamiltonian matrices on the RHS
scale as $\mathcal{O}(M^2 K_o^2 K_v^2)\sim \mathcal{O}(M^2 K^4)$, with $K_o$ and $K_v$ $(= K-K_o)$
being the numbers of occupied and virtual orbitals, respectively. Therefore,
IN-GMS-SU-CCSD is about $N_{it}\times\max(M,M^2/K^2)$ times more expensive than SR-CCSD, with $N_{it}$ being the number of macro-iterations required
to solve Eq. \eqref{SU-CC-T} self-consistently (NB: given an estimate of the RHS of Eq. \eqref{SU-CC-T}, solving iteratively for the SR-CCSD amplitudes
for each reference NED is considered as mirco-iterations). All in all, it is hardly possible to run IN-GMS-SU-CCSD even for a GMS composed only of
a few hundreds of reference NEDs. For this reason, it is necessary to formulate a MRPT2 variant of IN-GMS-SU-CCSD (see Sec. \ref{SecPASPT2}).

\subsection{PASPT2}\label{SecPASPT2}
Since the wave operator in Eq. \eqref{Omega} takes each reference NED $|\alpha\rangle$ as a distinct Fermi vacuum,
it is natural to introduce a reference-specific partitioning of the full Hamiltonian $H$, viz.,
\begin{align}
H=H_0^{\alpha}+V^{\alpha},
 \end{align}
where the zero-order term is subject to the condition [NB: $H_0^\alpha P_\beta|_{\beta\ne\alpha}$ is never needed]
\begin{align}
QH_0^\alpha P_\alpha= P_\alpha H_0^{\alpha\dag} Q=0, \label{H0property}
\end{align}
which implies that
\begin{align}
&H_0^\alpha P_\alpha= P H_0^\alpha P_\alpha, \nonumber\\
&QH_0^\alpha=\sum_{\beta\ne\alpha} QH_0^\alpha P_\beta+ QH_0^\alpha Q.\label{H0property1}
\end{align}

\begin{figure}[htbp]
\centering
\scalebox{0.6}{\includegraphics{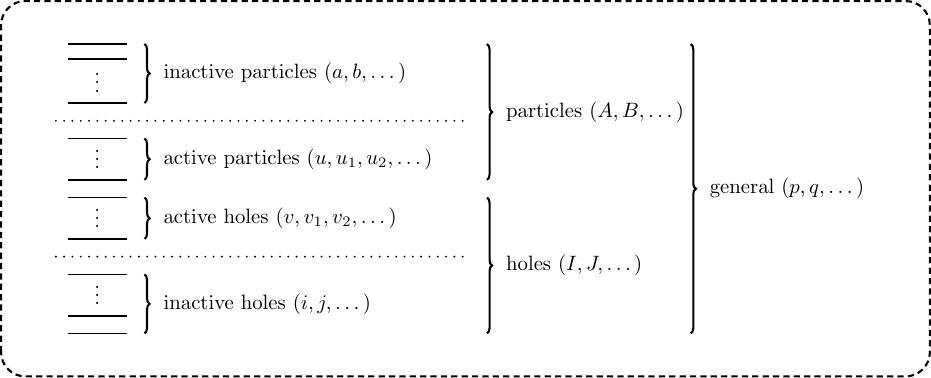}}
\caption{Classification of molecular spin orbitals with respect to the reference NED $|\alpha\rangle$.}
\label{fig:notations}
\end{figure}

\begin{figure}[htbp]
\centering
\scalebox{0.5}{\includegraphics{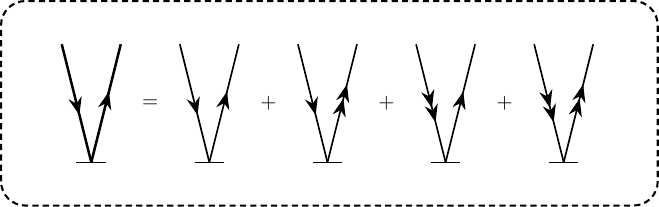}}
\caption{Diagrammatic representation of one-body cluster operators with $|\alpha\rangle$ as the Fermi vacuum.
Thick lines with up/down-going arrows denote general particles/holes; thin lines with up/down-going double (single) arrows
denote active (inactive) particles/holes.}
\label{One-T}
\end{figure}

\begin{figure}[htbp]
\centering
\scalebox{0.5}{\includegraphics{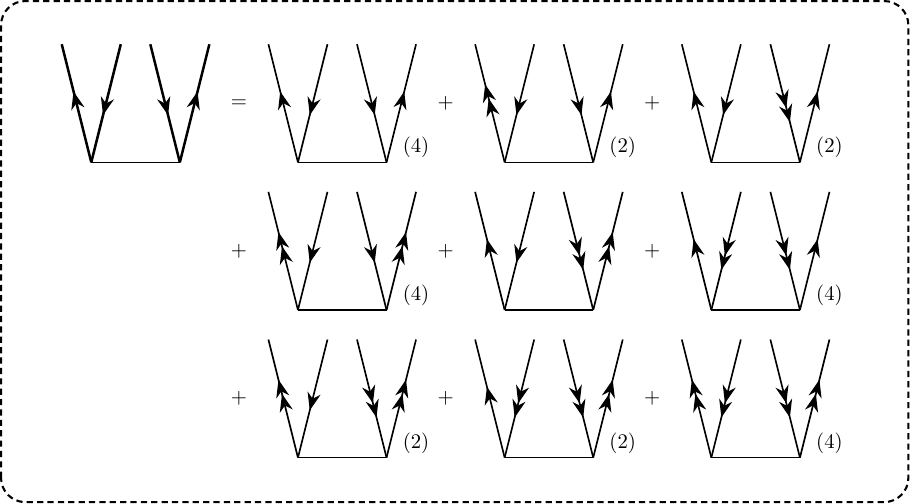}}
\caption{Diagrammatic representation of two-body cluster operators with $|\alpha\rangle$ as the Fermi vacuum.
Thick lines with up/down-going arrows denote general particles/holes; thin lines with up/down-going double (single) arrows
denote active (inactive) particles/holes.
In parentheses are the degenerate manifolds of the diagrams.}
\label{Two-T}
\end{figure}

The first order of Eq. \eqref{INC} reads
\begin{align}
\langle\beta|T^\alpha|\alpha\rangle=0\Leftrightarrow P T^\alpha P_\alpha=0,\quad T^\alpha P_\alpha=QT^\alpha P_\alpha,\label{1stC-Cond}
 \end{align}
which is just the first-order C-condition. However, this does \emph{not} imply $PT^\beta|\alpha\rangle=0$ as in the case of CAS.
Note that, to avoid over notations, we have dropped off the designation of first-order `(1)' in, e.g.,
the $T^{\alpha(1)}$-cluster operator.
Under the C-condition \eqref{1stC-Cond}, the internal excitations in the first-order $T_1^{\alpha}$- and $T_2^{\alpha}$-cluster operators should
be eliminated, thereby leading to
\begin{align}
T^{\alpha}_1&=\tilde{t}^I_A(\alpha)\{a^A_I\}_\alpha\nonumber\\
&=t_a^i(\alpha) \{a^a_i\}_\alpha + t_u^i(\alpha) \{a^u_i\}_\alpha + t_a^v(\alpha)\{a^a_v\}_\alpha+ \tilde{t}_{u}^v \{a^u_v\}_\alpha,\label{T1-Clusters}\\
T^{\alpha}_2&=\frac{1}{4}\tilde{t}_{AB}^{IJ}(\alpha)\{a^{AB}_{IJ}\}_\alpha\nonumber\\
&=\frac{1}{4} t_{ab}^{ij}(\alpha)\{a^{ab}_{ij}\}_\alpha + \frac{1}{2} t_{au}^{ij}(\alpha) \{a^{au}_{ij}\}_\alpha
+\frac{1}{4} t_{u_1u_2}^{ij}(\alpha)\{a^{u_1u_2}_{ij}\}_\alpha \nonumber\\
&+\frac{1}{2}t_{ab}^{iv}\{a^{ab}_{iv}\}_\alpha +\frac{1}{4}t_{ab}^{v_1v_2}\{a^{ab}_{v_1v_2}\}_\alpha+t_{au}^{iv}(\alpha)\{a^{au}_{iv}\}_\alpha\nonumber\\
&+\frac{1}{2} t_{au}^{v_1v_2}(\alpha)\{a^{au}_{v_1v_2}\}_\alpha + \frac{1}{2} t_{u_1u_2}^{iv}(\alpha)\{a^{u_1u_2}_{iv}\}_\alpha
+\frac{1}{4}\tilde{t}_{u_1u_2}^{v_1v_2}\{a_{v_1v_2}^{u_1u_2}\}_\alpha,\label{T2-Clusters}
\end{align}
where $\{I, J,\cdots\}$ and $\{A, B, \cdots \}$ denote the occupied and unoccupied spin orbitals in the Fermi vacuum $|\alpha\rangle$.
The former are further split into inactive holes $\{i,j,\cdots\}$ and active holes $\{v, v_1, v_2,\cdots \}$, whereas the latter
are further split into active particles $\{u, u_1, u_2,\cdots\}$ and inactive particles $\{a, b, \cdots\}$, see Fig. \ref{fig:notations} for an illustration.
The active holes and particles are also called collectively ``local active spin orbitals'' (LASO)\cite{GMS-SU-CC2003}
that distinguish $|\alpha\rangle$ from another reference NED $|\beta\rangle$, whereas the inactive holes and particles
are common to $|\alpha\rangle$ and $|\beta\rangle$. General spin orbitals are to be denoted as $\{p,q, \cdots\}$.
The tilde over $t$ emphasizes the exclusion of internal excitations.
Under such conventions, the one- and two-body cluster operators can be
represented by the diagrams in Figs. \ref{One-T} and \ref{Two-T}, respectively.

\begin{figure}[htbp]
	\centering
	\includegraphics[width=0.5\textwidth]{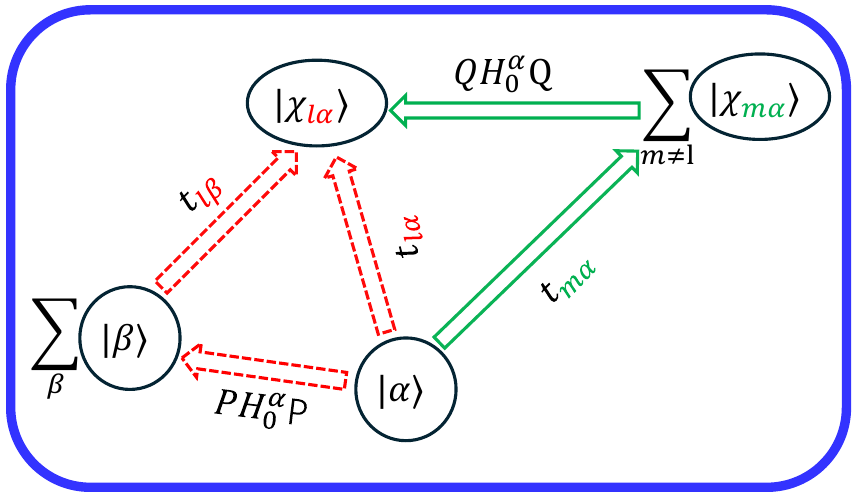}
	\caption{$Q$- and $P$-space couplings to the $t_{l\alpha}$-amplitudes of external excitations $\{\chi_{l\alpha}\}$ from the reference NED $|\alpha\rangle$
(i.e., processes for the dynamic correlation correction to $|\alpha\rangle$). }
	\label{t-couplings}
\end{figure}

The first order of Eq. \eqref{SU-CC-T} takes the following form
\begin{align}
\langle\chi_{l\alpha}|V^\alpha+[H_0^\alpha,T^\alpha]|\alpha\rangle = \sum_{\beta \ne\alpha\in \mathcal{M}_0} C_{l\alpha,\beta}H^0_{\beta\alpha},\label{1stEq}
\end{align}
where
\begin{align}
C_{l\alpha,\beta}&=\langle\chi_{l\alpha}|T^\beta-T^\alpha|\beta\rangle, \label{1stCoup}\\
H^0_{\beta\alpha}&=\langle\beta|H_0^\alpha|\alpha\rangle.
 \end{align}
Expanding the commutator in Eq. \eqref{1stEq} leads to
\begin{align}
&V_{l\alpha}+\langle\chi_{l\alpha}|QH_0^\alpha QT^\alpha|\alpha\rangle - t_{l\alpha} H^0_{\alpha\alpha}
= \sum_{\beta \ne\alpha\in \mathcal{M}_0} t_{l\beta} H^0_{\beta\alpha},\label{MP1a}\\
&V_{l\alpha}=\langle\chi_{l \alpha}|V^\alpha|\alpha\rangle,\\
&t_{l \beta}=\langle\chi_{l \alpha}|T^\beta|\beta\rangle.
\end{align}
It follows that the terms in Eq. \eqref{1stEq} that involve the amplitude $\langle\chi_{l\alpha}|T^\alpha|\beta\rangle$
for the transition from the reference NED $|\beta\rangle$ to the same external function $|\chi_{l\alpha}\rangle$ as from the reference NED $|\alpha\rangle$
are canceled out, meaning that, at the first order, a cluster operator is effective only when acting on its own Fermi vacuum.
It is also of interest to see that Eq. \eqref{1stEq} will reduce to the standard Rayleigh-Schr\"odinger type of SS-MRPT2
if the dimension of $\mathcal{M}_0$ is reduced to one and meanwhile the reference $|\alpha\rangle$ is chosen to
be an eigenfunction of $H_0^\alpha$ (which is usually a CAS state as in NEVPT2\cite{NEVPT2a} and CASPT2\cite{CASPT2b}).
Eq. \eqref{MP1a} can further be written as
\begin{align}
t_{l\alpha}&=-\frac{V_{l\alpha}}{\Delta E_{l\alpha}}
-\frac{\sum_{m\ne l\in Q} H^0_{l\alpha,m\alpha} t_{m \alpha}}
{\Delta E_{l\alpha}}
+\frac{\sum_{\beta\ne \alpha \in \mathcal{M}_0} t_{l \beta} H^0_{\beta\alpha}}
{\Delta E_{l\alpha}},\label{MP1}\\
H^0_{l\alpha,m\alpha}&=\langle\chi_{l \alpha}|QH_0^\alpha Q|\chi_{m \alpha}\rangle,\label{QH0Qmat}\\
\Delta E_{l\alpha}&=H^0_{l\alpha,l\alpha}-E_{\alpha}^0,\quad E_\alpha^0=H^0_{\alpha\alpha}.
\end{align}
It can then be seen that the first term of Eq. \eqref{MP1} is just the standard expression for the first-order $\mathbf{t}$-amplitudes
of individual reference NEDs, whereas the second and third terms represent their couplings, between
different $\mathbf{t}$-amplitudes of the same reference NED and between those of different reference NEDs, respectively (cf. Fig. \ref{t-couplings}).
For the purpose of implementation, Eq. \eqref{MP1} can be recast into an inhomogeneous system of linear equations
\begin{align}
\mathbf{A}\mathbf{t}=\mathbf{V},\label{LinearSys}\\
A_{l\alpha,m\alpha}&=\langle\chi_{l\alpha}|E_\alpha^0-H_0^\alpha|\chi_{m\alpha}\rangle,\label{A_matrix_elmt_1}\\
 A_{l\alpha,l\beta}&=H^0_{\beta\alpha}, \quad \beta\ne\alpha\in\mathcal{M}_0,\label{A_matrix_elmt_2}
\end{align}
where both $\{t_{l\alpha}\}$ and $\{V_{l\alpha}\}$ are understood as column vectors, whereas $\mathbf{A}$ is a sparse matrix
[the number of nonvanishing elements scales as $\mathcal{O}(M(K_o+K_v)K_o^2K_v^2)\sim \mathcal{O}(MK^5)$].
It is obvious that the number of unknown amplitudes $\{t_{l\alpha}\}$ is equal to that of the determining conditions
$(\mathbf{At})_{l\alpha}=V_{l\alpha}$, so that Eq. \eqref{LinearSys} has a unique solution, provided that $\mathbf{A}$ is nonsingular.

\begin{figure}[H]
	\centering
\begin{threeparttable}
		\begin{tabular}{cc}
			\includegraphics[width=0.6\linewidth]{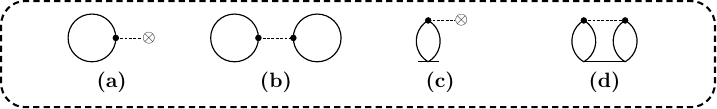} & \\
(a) \\ \\
			\includegraphics[width=0.6\linewidth]{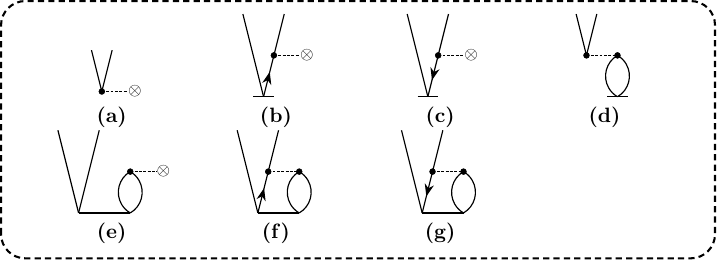} &  \\
(b) \\ \\
			\includegraphics[width=0.6\linewidth]{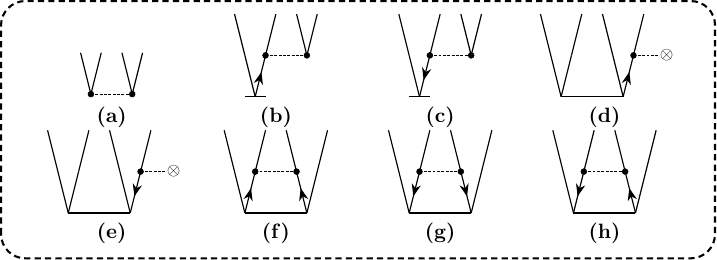} &  \\
(c) \\ \\
            \includegraphics[width=0.6\linewidth]{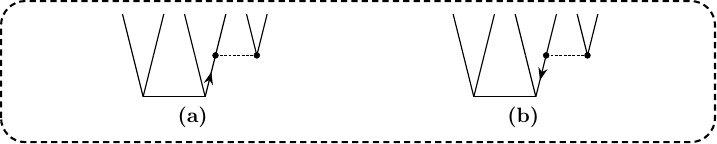} &  \\
(d) \\ \\
		\end{tabular}
\end{threeparttable}
\caption{Diagrammatic representation of the normal-ordered, up to three-body effective Hamiltonian $\bar{H}^{\mathrm{eff}[2]}$
in Eq. \eqref{PASH2nd} with $|\alpha\rangle$ as the Fermi vacuum. (a) $\bar{H}^{\mathrm{eff}[2]}_0$;
(b) $\bar{H}^{\mathrm{eff}[2]}_1$; (c) $\bar{H}^{\mathrm{eff}[2]}_2$; (d) $\bar{H}^{\mathrm{eff}[2]}_3$.  }
\label{fig:Heff}
\end{figure}

Given the (connected) $\mathbf{t}$-amplitudes (see Sec. \ref{Extensivity}), the up-to second-order effective Hamiltonian can be
constructed as
\begin{align}
H^{\mathrm{eff}[2]}&=P[H + \sum_{\alpha}  H T^\alpha P_\alpha]P\label{PASH2a}\\
&= P H P + \sum_{\alpha}  \sum_{l \in Q} t_{l\alpha} P H |\chi_{l\alpha}\rangle\langle\alpha|.
\end{align}
However, the second term may contain disconnected terms. For instance, if $|\chi_{l\alpha}\rangle=a_{v_1}^{u_1}|\alpha\rangle$ and
$|\beta\rangle=a_{v_1v_2}^{u_1u_2}|\alpha\rangle=a_{v_2}^{u_2}|\chi_{l\alpha}\rangle$,
then the matrix element $\langle \beta|H |\chi_{l\alpha}\rangle$ [$=f_{u_2}^{v_2}(\alpha)$] will depend only on the indices $u_2$ and $v_2$ and is hence not connected with
the amplitude $t_{l\alpha}=t^{v_1}_{u_1}(\alpha)$. To avoid this, Eq. \eqref{PASH2a} should be modified to
\begin{align}
\bar{H}^{\mathrm{eff}[2]} &=P[H + \sum_\alpha \acontraction[0.5ex]{}{H}{ }{T^\alpha} H T^\alpha P_\alpha]P, \label{PASH2}
\end{align}
where $\acontraction[0.5ex]{}{H}{ }{T^\alpha} H T^\alpha=[H, T^\alpha]$, emphasizing the connectivity, which is manifested more clearly
by normal ordering $\bar{H}^{\mathrm{eff}[2]}$ with respect to $|\alpha\rangle$, viz.
\begin{align}
\bar{H}^{\mathrm{eff}[2]}&=\bar{H}^{\mathrm{eff}[2]}_0+\bar{H}^{\mathrm{eff}[2]}_1+\bar{H}^{\mathrm{eff}[2]}_2+\bar{H}^{\mathrm{eff}[2]}_3, \label{PASH2nd}
\end{align}
where
%
%
%
%
\begin{align}
\bar{H}^{\mathrm{eff}[2]}_0&=\langle\alpha|H|\alpha\rangle+\langle\alpha|[H,T^\alpha]|\alpha\rangle=E^{[1]}_{\alpha}+E^{(2)}_{\alpha}\label{MP2E}\\
&=\left[h_I^I+\frac{1}{2}\bar{g}_{IJ}^{IJ}\right]+\left[f_I^A(\alpha)t_A^I(\alpha)+\frac{1}{4}\bar{g}_{IJ}^{AB}t_{AB}^{IJ}(\alpha)\right],\\
\bar{H}^{\mathrm{eff}[2]}_1&=\left[f_A^I(\alpha)+f_A^B(\alpha)t_B^I(\alpha)-f_J^I(\alpha)t_A^J(\alpha)
+\bar{g}_{AJ}^{IB}t_B^J(\alpha)\right.\nonumber\\
&\left. +f_J^B(\alpha)t_{AB}^{IJ}(\alpha)+\frac{1}{2}\bar{g}_{AJ}^{CB}t_{CB}^{IJ}(\alpha)-\frac{1}{2}\bar{g}_{KJ}^{IB}t_{AB}^{KJ}(\alpha)\right]\{a^{A}_{I}\}_\alpha,\\
\bar{H}^{\mathrm{eff}[2]}_2&=\left[\frac{1}{4}\bar{g}_{AB}^{IJ}+\frac{1}{2}\bar{g}_{AB}^{CJ}t_C^I(\alpha)
-\frac{1}{2}\bar{g}_{AK}^{IJ}t_B^K(\alpha)
+\frac{1}{2}f_B^C(\alpha)t_{AC}^{IJ}(\alpha)-\frac{1}{2}f_K^J(\alpha)t_{AB}^{IK}(\alpha)\right.\nonumber\\
&\left. +\frac{1}{8}\bar{g}_{AB}^{CD}t_{CD}^{IJ}(\alpha)+\frac{1}{8}\bar{g}_{KL}^{IJ}t_{AB}^{KL}(\alpha)
-\bar{g}_{AK}^{CJ}t_{CB}^{IK}(\alpha)\right]\{a^{AB}_{IJ}\}_\alpha,\\
\bar{H}^{\mathrm{eff}[2]}_3&=\left[\frac{1}{4}\bar{g}_{AB}^{DJ}t_{DC}^{IK}(\alpha)-\frac{1}{4}\bar{g}_{LC}^{IK}t_{AB}^{LJ}(\alpha)\right]\{a^{ABC}_{IJK}\}_\alpha.
\end{align}
The diagrammatic representations of $\bar{H}^{\mathrm{eff}[2]}_i$ ($i\in[0,3]$)
are shown in Fig. \ref{fig:Heff}. It is seen that every diagram is connected.

However, the effective Hamiltonian $\bar{H}^{\mathrm{eff}[2]}$ \eqref{PASH2}/\eqref{PASH2nd} may not be a closed operator for the chosen PAS $\mathcal{M}_0$.
That is, some of the second quantized excitation operators in $\bar{H}^{\mathrm{eff}[2]}$ \eqref{PASH2nd} may be quasi-open,
so that the energies $E_I^{[2]}$ obtained by diagonalization of $\bar{H}^{\mathrm{eff}[2]}$ are in general disconnected\cite{MukerjeeClosedOp}.
This can be resolved in two different ways. One is to set\cite{MukerjeeClosedOp} the matrix elements $\bar{H}^{\mathrm{eff}[2]}_{\beta\alpha}$ to zero, provided that
\begin{align}
A|\alpha\rangle=|\beta\rangle,\quad A|\gamma\rangle= |\delta\rangle \in \mathcal{M}_0^\perp,
\quad |\alpha\rangle, |\beta\rangle,  |\gamma\rangle\in\mathcal{M}_0,\label{Q-open}
\end{align}
where $A$ represents one of the active excitation operators $\{a_{v}^{u}\}_\alpha$,  $\{a_{v_1v_2}^{u_1u_2}\}_\alpha$
and $\{a_{v_1v_2v_3}^{u_1u_2u_3}\}_\alpha$ in $\bar{H}^{\mathrm{eff}[2]}$ and is quasi-open
(i.e., $A$ can connect two reference NEDs but also can excite at least one reference NEDs to $\mathcal{M}_0^\perp$)\cite{MukerjeeClosedOp}.
The so-polished effective Hamiltonian is to be denoted as $\bar{H}^{\mathrm{eff}[2]}_Q$.
The other is to bring $A|\gamma\rangle$ into $\mathcal{M}_0$, so as to obtain an
extended model space $\mathcal{M}_X$. The procedure is repeated for all reference NEDs and the
active excitation operators in $\bar{H}^{\mathrm{eff}[2]}$, until that $\bar{H}^{\mathrm{eff}[2]}$ becomes just closed.
Note that the so-constructed $\mathcal{M}_X$ is in general still much smaller than the corresponding CAS $\mathcal{M}_C$
(i.e., $\mathcal{M}_X\bigcup\bar{\mathcal{M}}_X=\mathcal{M}_C$).
In practice, the correlation correction to the newly added reference NEDs can be ignored.
In this case, the matrix elements of the \emph{intermediate} Hamiltonian (denoted as $\bar{H}^{\mathrm{eff}[2]}_X$) read
\begin{align}
(\bar{H}^{\mathrm{eff}[2]}_X)_{\beta\alpha}&=
\begin{cases} \bar{H}^{\mathrm{eff}[2]}_{\beta\alpha},\quad \beta \mbox{ and } \alpha\in\mathcal{M}_0,\\
H_{\beta\alpha},\quad \beta  \mbox{ or } \alpha\in\bar{\mathcal{M}}_0=\mathcal{M}_X\backslash\mathcal{M}_0.
\end{cases}\label{HeffX}
\end{align}
This amounts to treating the newly added reference NEDs spanning the space $\bar{\mathcal{M}}_0$ as buffers.
That is, the $T^\alpha$-clusters are subject to the following extended C-conditions
\begin{align}
\langle \beta | T^\alpha |\alpha\rangle& = 0, \quad \forall \beta\in \mathcal{M}_X, \quad \alpha\in \mathcal{M}_0,\\
T^\beta|\beta\rangle&=0,\quad \forall \beta\in \bar{\mathcal{M}}_0=\mathcal{M}_X\backslash\mathcal{M}_0,
\end{align}
which dictate that only those functions belonging to $\mathcal{M}_0$ are allowed to excite to $\mathcal{M}_X^\perp$ ($=\bar{\mathcal{M}}_X+\mathcal{M}_C^\perp$).
This has a sound physical basis: by construction, the functions belonging to $\bar{\mathcal{M}}_0$ do not contribute significantly to the desired states
but are prone to intruder states, so that they should not be perturbed.

After having determined the $\mathbf{t}$-amplitudes for the prechosen model space $\mathcal{M}_0$, we will obtain three sets of
(relaxed) energies by diagonalization of $\bar{H}^{\mathrm{eff}[2]}$, $\bar{H}^{\mathrm{eff}[2]}_Q$ and
$\bar{H}^{\mathrm{eff}[2]}_X$, respectively. It is expected that $\bar{H}^{\mathrm{eff}[2]}_X$
is more accurate than $\bar{H}^{\mathrm{eff}[2]}_Q$ because of its extended dimension as well as reduced nonhermiticity.
As a check of internal consistency,
the relaxed energies $E_I^{[2]}$ can be compared to the unrelaxed energies $\bar{E}_I^{[2]}$ calculated as
\begin{align}
\bar{E}_I^{[2]}&=\mathbf{C}^{(0)\dag}_I\mathbf{O}\mathbf{C}^{(0)}_I, \quad
\mathbf{O}=\bar{\mathbf{H}}^{\mathrm{eff}[2]}, \bar{\mathbf{H}}^{\mathrm{eff}[2]}_Q, \bar{\mathbf{H}}^{\mathrm{eff}[2]}_X, \label{unrelaxedE}
\end{align}
where $\mathbf{C}^{(0)}_I$ are the CI coefficients obtained by diagonalization of $PHP$.
The small differences $|\bar{E}_I^{[2]}-E_I^{[2]}|$ are a strong indicator for the good accuracy.
As a matter of fact, the diagonal elements of $\bar{H}^{\mathrm{eff}[2]}$ \eqref{PASH2}
already tell the appropriateness of the calculations: the first and second terms therein
are just the HF and
MP2 correlation energies, respectively [see $E_{\alpha}^{[1]}$ and $E_{\alpha}^{(2)}$ in Eq. \eqref{MP2E}].
If $E_{\alpha}^{(2)}$ is beyond some portion of $E_{\alpha}^{[1]}$ (say, $>1\%$), the results
are likely problematic. Why this can happen is because the reference-specific Fock operator \eqref{Fspin}
is constructed with orthonormal orbitals that are common to all reference NEDs and hence does not
satisfy the Brillouin conditions $f_I^A(\alpha)=f_A^I(\alpha)=0$ for $|\alpha\rangle$.
As a result, some external excitations (especially singles) from $|\alpha\rangle$
may have too large amplitudes, thereby rendering $E_{\alpha}^{(2)}$ unrealistic.
In this case, such contributions should be reduced (by shifting
the energy denominators $\Delta E_{l\alpha}$) or simply removed (NB: removing some but not all external functions $\chi_{l\alpha}$ does not affect
the connectedness of the $\mathbf{t}$-amplitudes). Note that this should not be confused with the intruder-state problem,
which is already resolved by the introduction of the buffer space $\bar{\mathcal{M}}_0$.


\subsubsection{Size-extensivity}\label{Extensivity}
\begin{figure}[htbp]
\centering
\setlength{\tabcolsep}{-2pt}
\scalebox{1.2}{\includegraphics{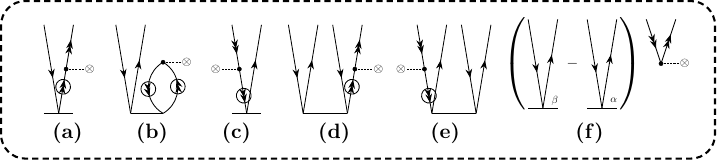}}
\caption{Diagrammatic representation of the one-body $P$-space couplings with $|\beta\rangle$ as the Fermi vacuum.
(a) to (f) correspond to Eqs. \eqref{Good1} to \eqref{Badcase}, respectively.
Double arrows enclosed by a circle indicate a change of
the Fermi vacuum, thereby carrying on an extra negative sign. }
\label{One-Coupling}
\end{figure}

Size-extensivity, i.e., connectedness of the energy\cite{Size-extensivity}, is a necessary condition for
a method to perform uniformly when the system size ever increases. To achieve this for a multi-reference method,
both the $\mathbf{t}$-amplitudes and the effective/intermediate Hamiltonian have to be connected.
Moreover, the effective/intermediate Hamiltonian has to be a closed operator\cite{MukerjeeClosedOp}.
To examine the connectedness of the RHS of Eq. \eqref{1stEq}, we take a generic zeroth-order Hamiltonian $H_0^\alpha$
\begin{align}
H_0^\alpha&=f^\alpha+g,\label{H0def}
\end{align}
where $f^\alpha$ is an effective one-body operator, whereas $g$ is a genuine two-body operator.
Since $H_0^\alpha$ cannot connect two reference NEDs that differ by more than two LASOs, only
$|\beta\rangle = a^u_v|\alpha\rangle$ and $|\beta\rangle = a^{u_1 u_2}_{v_1 v_2}|\alpha\rangle$
need to be discussed explicitly. For the former, there exist six types of external singles and doubles from $|\alpha\rangle$:
\begin{subequations}\label{1LASO}
\begin{align}
&|\chi_{l\alpha}\rangle = a^u_i |\alpha\rangle:\nonumber\\
& \sum_{\beta\ne\alpha}C_{l\alpha,\beta}H^0_{\beta\alpha}
= - \sum_{\beta\ne\alpha}f_u^v(\alpha)t_v^i(\beta), \label{Good1}
\end{align}
\begin{align}
&|\chi_{l\alpha}\rangle = a^a_i|\alpha\rangle:\nonumber\\
&\sum_{\beta\ne\alpha}C_{l\alpha,\beta}H^0_{\beta\alpha}
= \sum_{\beta\ne\alpha}f_u^v(\alpha)t_{av}^{iu}(\beta), \label{Good2}
\end{align}
\begin{align}
&|\chi_{l\alpha}\rangle = a^a_v|\alpha\rangle:\nonumber\\
&\sum_{\beta\ne\alpha}C_{l\alpha,\beta}H^0_{\beta\alpha}
= \sum_{\beta\ne\alpha}f_u^v(\alpha)t_a^u(\beta),  \label{Good3}
\end{align}
\begin{align}
&|\chi_{l\alpha}\rangle = a^{au}_{ij}|\alpha\rangle:\nonumber\\
&\sum_{\beta\ne\alpha}C_{l\alpha,\beta}H^0_{\beta\alpha}
= - \sum_{\beta\ne\alpha}f_u^v(\alpha)t_{av}^{ij}(\beta),  \label{Good4}
\end{align}
\begin{align}
&|\chi_{l\alpha} \rangle = a^{ab}_{iv}|\alpha\rangle:\nonumber\\
&\sum_{\beta\ne\alpha}C_{l\alpha,\beta}H^0_{\beta\alpha}
= \sum_{\beta\ne\alpha}f_u^v(\alpha)t_{ab}^{iu}(\beta), \label{Good5}
\end{align}
\begin{align}
&|\chi_{l\alpha}\rangle = a^{au}_{iv}|\alpha\rangle:\nonumber\\
&\sum_{\beta\ne\alpha}C_{l\alpha,\beta}H^0_{\beta\alpha}
= f_u^v(\alpha)[t_a^i(\beta) - t_a^i(\alpha)].\label{Badcase}
\end{align}
\end{subequations}
where $f_p^q(\alpha)=\langle p|f^\alpha|q\rangle$.
The diagrammatic representation of Eq. \eqref{1LASO} is shown in Fig. \eqref{One-Coupling}.

\begin{figure}[htbp]
\centering
\setlength{\tabcolsep}{-2pt}
\scalebox{1.2}{\includegraphics{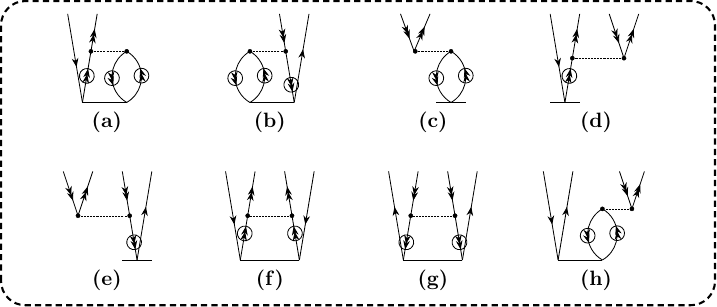}}
\caption{Diagrammatic representation of the two-body $P$-space couplings with $|\beta\rangle$ as the Fermi vacuum.
(a) to (h) correspond to Eqs. \eqref{2LASOa} to \eqref{2LASOh}, respectively.
Double arrows enclosed by a circle indicate that this index undergoes a change of
the Fermi vacuum, thereby carrying on an extra negative sign.}
\label{Two-Coupling}
\end{figure}

As for the case of $|\beta\rangle = a^{u_1 u_2}_{v_1 v_2}|\alpha\rangle$ with $u_1> u_2$ and $v_1> v_2$,
there are twenty types of external singles and doubles from $|\alpha\rangle$ [NB: $(m,\bar{m})=(1,2), (2,1))$]:
\begin{subequations}\label{2LASO}
\begin{align}
&|\chi_{l\alpha} \rangle = a^{u_m}_i| \alpha\rangle:\nonumber\\
&\sum_{\beta\ne\alpha}C_{l\alpha,\beta}H^0_{\beta\alpha}
= -\sum_{\beta\ne\alpha}\bar{g}_{u_m u_{\bar{m}}}^{v_m v_{\bar{m}}} t_{v_m v_{\bar{m}}}^{i u_{\bar{m}}}(\beta),\label{2LASOa}
\end{align}
\begin{align}
&|\chi_{l\alpha}\rangle = a^a_{v_m} |\alpha\rangle:\nonumber\\
&\sum_{\beta\ne\alpha}C_{l\alpha,\beta}H^0_{\beta\alpha}
= \sum_{\beta\ne\alpha}\bar{g}_{u_m u_{\bar{m}}}^{ v_m, v_{\bar{m}}} t_{a v_{\bar{m}}}^{u_m u_{\bar{m}}}(\beta),
\end{align}
\begin{align}
&|\chi_{l\alpha}\rangle = a^{u_m}_{v_n} |\alpha\rangle:\nonumber\\
&\sum_{\beta\ne\alpha}C_{l\alpha,\beta}H^0_{\beta\alpha}
= \sum_{\beta\ne\alpha}\bar{g}_{u_m u_{\bar{m}}}^{ v_n v_{\bar{n}}} t_{v_{\bar{n}}}^{u_{\bar{m}}}(\beta), \label{CAS0}
\end{align}
\begin{align}
&|\chi_{l\alpha}\rangle = a^{u_m u_{\bar{m}}}_{i v_{\bar{m}}}|\alpha\rangle:\nonumber\\
&\sum_{\beta\ne\alpha}C_{l\alpha,\beta}H^0_{\beta\alpha}
= - \sum_{\beta\ne\alpha}\bar{g}_{u_m u_{\bar{m}}}^{v_m v_{\bar{m}}} t_{v_m}^i(\beta),
\end{align}
\begin{align}
&|\chi_{l\alpha}\rangle = a^{u_m a}_{v_m v_{\bar{m}}}|\alpha\rangle:\nonumber\\
&\sum_{\beta\ne\alpha}C_{l\alpha,\beta}H^0_{\beta\alpha}
= \sum_{\beta\ne\alpha}\bar{g}_{u_m u_{\bar{m}}}^{v_m v_{\bar{m}}} t_a^{u_{\bar{m}}}(\beta),
\end{align}
\begin{align}
&|\chi_{l\alpha}\rangle = a^{u_m u_{\bar{m}}}_{i j}|\alpha\rangle:\nonumber\\
&\sum_{\beta\ne\alpha}C_{l\alpha,\beta}H^0_{\beta\alpha}
= \sum_{\beta\ne\alpha}\bar{g}_{u_m u_{\bar{m}}}^{v_m v_{\bar{m}}} t_{v_m v_{\bar{m}}}^{i j}(\beta),
\end{align}
\begin{align}
&|\chi_{l\alpha}\rangle = a^{a b}_{v_m v_{\bar{m}}}|\alpha\rangle:\nonumber\\
&\sum_{\beta\ne\alpha}C_{l\alpha,\beta}H^0_{\beta\alpha}
= \sum_{\beta\ne\alpha}\bar{g}_{u_m u_{\bar{m}}}^{v_m v_{\bar{m}}} t_{a b}^{u_m u_{\bar{m}}}(\beta),
\end{align}
\begin{align}
&|\chi_{l\alpha}\rangle = a^{a u_m}_{i v_n} |\alpha\rangle:\nonumber\\
&\sum_{\beta\ne\alpha}C_{l\alpha,\beta}H^0_{\beta\alpha}
= \sum_{\beta\ne\alpha} \bar{g}_{u_m u_{\bar{m}}}^{v_n v_{\bar{n}}} t_{a v_{\bar{n}}}^{i u_{\bar{m}}}(\beta),\label{2LASOh}
\end{align}
\end{subequations}	
where $\bar{g}_{pq}^{rs}=(pr|qs)-(ps|qr)$ in the Mulliken notation.
The diagrammatic representation of Eq. \eqref{2LASO} is shown in Fig. \eqref{Two-Coupling}.

It is easily seen that all the products in Eq. \eqref{1LASO} [or Fig. \eqref{One-Coupling}] and Eq. \eqref{2LASO} [or Fig. \eqref{Two-Coupling}]
have common indices and are hence connected,
with the exception of that in Eq. \eqref{Badcase},
which is possibly connected only when
$a_i^a|\alpha\rangle$ is an external function (since $a_i^a|\beta\rangle$ is an external function).
However, this is not true in general for an arbitrary PAS. That is, the RHS of eq (60f) is in general
disconnected.
The same situation occurs also to the IN-GMS-SU-CCSD amplitude equations
(see the first term on the RHS of equation (98) in Ref. \citenum{GMS-SU-CC2003}, which is to be multiplied with $H^{\mathrm{eff}}_{\beta\alpha}$
in place of the present $H^0_{\beta\alpha}$).
Therefore, IN-GMS-SU-CC is \emph{not} size-extensive at all, contrary to their claim\cite{GMS-SU-CC2003,PaldusJMC2004}
(see also the formal analysis by Nooijen and coworkers\cite{nooijen2005reflections}).
On the other hand, it can be proven\cite{JM1981} that the common-orbital terms of the $\mathbf{t}$-amplitudes
of different references [e.g., $t^i_a(\beta)$ and $t^i_a(\alpha)$ in Eq. \eqref{Badcase}] are identical and hence
cancel each other in the case of CAS, so that IN-CAS-SU-CC\cite{JM1981} is indeed size-extensive.
As for PASPT2, it is clear that Eqs. \eqref{1LASO} and \eqref{2LASO} impose restrictions only on the off-diagonal elements
of $P H_0^\alpha P^\alpha$, which suggests the following form
\begin{align}
H_{0,\mathrm{A}}^{\alpha}&=\sum_{\beta, \alpha(\ne\beta)}^\prime P_\beta H P_\alpha
\end{align}
for the active part $H_{0,\mathrm{A}}^{\alpha}$ of $H_0^\alpha$,
\begin{align}
H_0^\alpha&=H_{0,\mathrm{I}}^{\alpha}+H_{0,\mathrm{A}}^{\alpha}.\label{H0final}
\end{align}
However, a second-quantized form of $H_{0,\mathrm{A}}^{\alpha}$ must be adopted, which drives us to define it as
\begin{align}
H_{0,\mathrm{A}}^{\alpha}&=\sum_{\beta, \alpha(\ne\beta)}^\prime \langle\beta|f^\alpha|\alpha\rangle S_{\beta\alpha}
+\sum_{\beta, \alpha(\ne\beta)}\langle\beta|g|\alpha\rangle D_{\beta\alpha}  \label{H0ActiveA}\\
&\equiv\sum_{v,u}^{\prime} f_u^v(\alpha)\{a^u_v\}_{\alpha}
 +\frac{1}{2}\sum_{u_1,u_2,v_1,v_2}g_{u_1 u_2}^{v_1v_2}\{a^{u_1u_2}_{v_1v_2}\}_{\alpha},\label{H0Active}
\end{align}
where $S_{\beta\alpha}$ and $D_{\beta\alpha}$ are single
and double transfer operators, respectively (i.e., $|\beta\rangle=\Omega_{\beta\alpha}|\alpha\rangle$).
They are just \emph{internal} excitation operators $\{a^u_v\}_{\alpha}$
and $\{a^{u_1u_2}_{v_1v_2}\}_{\alpha}$, respectively.
The prime in the first summation of Eq. \eqref{H0ActiveA}/\eqref{H0Active} emphasizes the exclusion of the case of
$|\beta\rangle=a_v^u|\alpha\rangle$ \emph{and} $|\chi_{l\alpha}\rangle=a_{iv}^{au}|\alpha\rangle$,
amounts to setting  Eq. \eqref{Badcase} to zero. Further combined with the facts that
$[H_{0,\mathrm{A}}^{\alpha}, T^\alpha]=0$ and that Eqs. \eqref{Good1} to \eqref{Good5} do not involve $\sum_{\beta\ne\alpha}\langle\chi_{l\alpha}|T^\alpha|\beta\rangle\langle\beta|H_{0A}^\alpha|\alpha\rangle$ at all,
the $QH_0^\alpha Q$ operator in Eqs. \eqref{MP1a} and \eqref{QH0Qmat} must be replaced with $QH_{0I}^\alpha Q$.
Meanwhile, the $t_{l\beta}$ in Eqs. \eqref{MP1a} and \eqref{MP1} must also be set to zero for this particular pair
of $|\beta\rangle$ and $|\chi_{l\beta}\rangle$.

The inactive part $H_{0,\mathrm{I}}^{\alpha}$ of $H_0^\alpha$ can readily be deduced
from the fact that the LHS of Eq. \eqref{1stEq} has the same form
as that in the single-reference second-order M{\o}ller-Plesset perturbation theory (MP2)\cite{MP2}, viz.,
\begin{align}
H_{0,\mathrm{I}}^{\alpha}&= f_I^J(\alpha)a^I_J+ f_A^B(\alpha)a^A_B \nonumber\\
&=f_I^I(\alpha)+ f_I^J(\alpha)\{a^I_J\}_\alpha+ f_A^B(\alpha)\{a^A_B\}_\alpha.\label{H0Inactive}
\end{align}
It can readily be checked that $QH_0^\alpha P_\alpha=Q(H_{0,\mathrm{I}}^\alpha+H_{0,\mathrm{A}}^\alpha) P_\alpha=0$.
Since $[H_{0}^{\alpha}, T^\alpha]\equiv\acontraction[0.5ex]{}{H_{0,\mathrm{I}}^{\alpha}}{}{T^\alpha} H_{0,\mathrm{I}}^{\alpha} T^\alpha$,
the connectedness of the LHS of Eq. \eqref{1stEq} is obvious.
Therefore, Eq. \eqref{1stEq} is termwise connected with this particular definition of $H_0^\alpha$. So is
Eq. \eqref{MP1a}, as shown in Sec. \ref{SecCommutator}.

The above findings also hold for a CAS. It is just that Eq. \eqref{CAS0} is in this case always equal to zero
(because $a^{v_2}_{u_2}|\beta\rangle=a^{v_2}_{u_2}a_{v_1v_2}^{u_1u_2}|\alpha\rangle=a_{v_1}^{u_1}|\alpha\rangle$
is a reference NED, so that $t_{v_2}^{u_2}(\beta)=0$); the two terms in Eq. \eqref{Badcase} are truly connected\cite{JM1981},
provided that the following $H_0^{\mathrm{d}}$
 \begin{align}
 H_0^{\mathrm{d}}=\epsilon_p a^p_p\label{H0diag}
 \end{align}
is employed for the diagonal component of $H_0^\alpha$.
However, neither $H_0^\alpha=H_0^{\mathrm{d}}$ nor $H_0^\alpha=H_0^{\mathrm{d}}$ augmented with the off-diagonal
terms of $H_{0,\mathrm{I}}^{\alpha}$ \eqref{H0Inactive} is acceptable: the former
misses all couplings, whereas the latter may yield nonphysical (complex-valued) solutions even for very simple situations
(e.g., some vertical excitations of \ce{H2O} at the equilibrium geometry).
Therefore, even when working with a CAS, the use of the zeroth-order Hamiltonian defined in Eq. \eqref{H0final}
is a must.

Since the (orbital) class-diagonal operator $H_{0,\mathrm{I}}^{\alpha}$ \eqref{H0Inactive} cannot achieve spin adaptation (see Sec. \ref{SecCommutator}),
the Fock matrix $\mathbf{f}(\alpha)$ in Eqs. \eqref{H0Active} and \eqref{H0Inactive} is defined simply
in the spin orbital basis
\begin{equation}
f_p^q(\alpha)=h_p^q+\sum_{i}\bar{g}_{pi}^{qi}n_i(\alpha), \label{Fspin}
\end{equation}
although the same spatial orbital is always used for different spins.

\subsubsection{Internal consistency of $H_0^\alpha$ \eqref{H0final} }\label{SecCommutator}
The internal consistency of the peculiar $H_0^\alpha$ operator in Eq. \eqref{H0final}, along with Eqs. \eqref{H0Active} and \eqref{H0Inactive},
 requires a double check, especially
when going from Eq. \eqref{1stEq} to Eq. \eqref{MP1a}. To this end, we split the commutator on the LHS of Eq. \eqref{1stEq}
by virtue of the conditions in Eqs. \eqref{H0property} to \eqref{1stC-Cond},
\begin{align}
\langle\chi_{l\alpha}|[H_0^\alpha, T^\alpha]|\alpha\rangle&=\langle\chi_{l\alpha}|H_0^\alpha T^\alpha|\alpha\rangle-
\langle\chi_{l\alpha}|T^\alpha H_0^\alpha |\alpha\rangle\nonumber\\
&=\langle\chi_{l\alpha}|H_0^\alpha QT^\alpha|\alpha\rangle-
\langle\chi_{l\alpha}|T^\alpha P H_0^\alpha |\alpha\rangle\nonumber\\
&=\langle\chi_{l\alpha}|H_{0I}^\alpha QT^\alpha|\alpha\rangle
-\langle\chi_{l\alpha}|T^\alpha |\alpha\rangle H^0_{\alpha\alpha}  \nonumber\\
&+\langle\chi_{l\alpha}|H_{0A}^\alpha QT^\alpha|\alpha\rangle
-\sum_{\beta\ne\alpha} \langle\chi_{l\alpha}|T^\alpha |\beta\rangle H^0_{\beta\alpha},\label{SplitCommu}
\end{align}
where $H^0_{\beta\alpha}|_{\beta\ne\alpha}$ is by definition equal to $\langle\beta|H_{0,\mathrm{A}}^\alpha|\alpha\rangle$.
The last term of Eq. \eqref{SplitCommu}
is to cancel the same term on the RHS of Eq. \eqref{1stEq}, thereby leading to the RHS of Eq. \eqref{MP1a}.
Close inspections reveal that the third term in Eq. \eqref{SplitCommu} is nonvanishing only when
$|\chi_{l\alpha}=a_{iv}^{au}|\alpha\rangle$ and $T^\alpha|\alpha\rangle=a_i^a|\alpha\rangle t_a^i(\alpha)$,
so that $\langle\chi_{l\alpha}|H_{0A}^\alpha QT^\alpha|\alpha\rangle=f_u^v(\alpha)t_a^i(\alpha)$, which also appears
in Eq. \eqref{Badcase}. Since Eq. \eqref{Badcase} is set to zero by the restriction imposed on $H_{0A}^\alpha$ \eqref{H0Active},
the $QH_{0}^\alpha Q$ operator in Eqs. \eqref{MP1a} and \eqref{QH0Qmat} must be replaced with $QH_{0I}^\alpha Q$ for consistency.
This is also consistent with the fact that Eqs. \eqref{Good1} to \eqref{Good5} do not involve
$\sum_{\beta\ne\alpha}\langle\chi_{l\alpha}|T^\alpha|\beta\rangle\langle\beta|H_{0A}^\alpha|\alpha\rangle$ at all.
Note that, to be internally consistent, the $t_{l\beta}$ in Eqs. \eqref{MP1a} and \eqref{MP1} should also be set to zero for this particular pair
of $|\beta\rangle$ and $|\chi_{l\beta}\rangle$.

The first two terms of Eq. \eqref{SplitCommu} can be calculated as follows.
If $|\chi_{l\alpha}\rangle=a_I^A|\alpha\rangle$, then
\begin{align}
&\langle\chi_{l\alpha}|H_{0I}^\alpha|\chi_{m\alpha}\rangle t_{m\alpha}-t_{l\alpha}H^0_{\alpha\alpha}\nonumber\\
&=\left[f_K^L(\alpha) \langle\alpha|a_A^I a^K_L a^B_J|\alpha\rangle t_B^J(\alpha)-t_A^I(\alpha) H^0_{\alpha\alpha}\right]\nonumber\\
&+f_C^D(\alpha) \langle\alpha|a_A^I a^C_D a^B_J|\alpha\rangle t_B^J(\alpha)\nonumber\\
&=\left[-f_J^I(\alpha)t^J_A(\alpha)+t^I_A(\alpha)(f_K^K(\alpha)-H^0_{\alpha\alpha})\right]
+f_A^B(\alpha)t_B^J(\alpha)\nonumber\\
&=-f_J^I(\alpha)t^J_A(\alpha)+f_A^B(\alpha)t_B^I(\alpha),\label{SS}
\end{align}
where the second, disconnected term in the brackets vanishes only if $H^0_{\alpha\alpha}= f_K^K(\alpha)$,
which is nothing but $\langle\alpha|H_{0,\mathrm{I}}^\alpha|\alpha\rangle$.
On the other hand, if $|\chi_{l\alpha}\rangle=a_{IJ}^{AB}|\alpha\rangle$ with $A>B$ and $I>J$, then
\begin{align}
&\langle\chi_{l\alpha}|H_{0I}^\alpha|\chi_{m\alpha}\rangle t_{m\alpha}-t_{l\alpha}H^0_{\alpha\alpha}\nonumber\\
&=\left[f_M^N(\alpha) \sum_{C>D, K>L}\langle\alpha|a^{IJ}_{AB} a^M_N a^{CD}_{KL}|\alpha\rangle t^{KL}_{CD}(\alpha)-t_{AB}^{IJ}(\alpha) H^0_{\alpha\alpha}\right]\nonumber\\
&+f_F^E(\alpha) \sum_{C>D, K>L}\langle\alpha|a^{IJ}_{AB} a^E_F a^{CD}_{KL}|\alpha\rangle t^{KL}_{CD}(\alpha)\nonumber\\
&=-\mathcal{A}(IJ)f_K^I(\alpha)t^{KJ}_{AB}(\alpha)+\mathcal{A}(AB)f_A^C(\alpha)t^{IJ}_{CB}(\alpha),\label{DD}
\end{align}
where the antisymmetrizer reads $\mathcal{A}(IJ)=1-(IJ)$, with $(IJ)$ being the transposition of the indices $I$ and $J$.
Again, the cancellation of the second, disconnected term in the brackets requires that $H^0_{\alpha\alpha}=\langle\alpha|H_{0,\mathrm{I}}^\alpha|\alpha\rangle$.
Eqs. \eqref{SS} and \eqref{DD} describes the couplings between external singles and singles (SS) and between external doubles and doubles (DD), respectively.
Their diagonal terms can readily be extracted to be
$[f_A^A(\alpha)-f_I^I(\alpha)]t^I_A(\alpha)$ and $[f_A^A(\alpha)+f_B^B(\alpha)-f_I^I(\alpha)-f_J^J(\alpha)]t^{IJ}_{AB}(\alpha)$,
respectively. This completes the derivation Eq. \eqref{MP1a}, which is termwise connected as Eq. \eqref{1stEq}.

For completeness, we also document the corresponding expressions for the class-off-diagonal operator
\begin{align}
O^\alpha_{0,\mathrm{I}}=f_A^I(\alpha)a^A_I+f_I^A(\alpha)a^I_A.
\end{align}
If $|\chi_{l\alpha}\rangle=a_I^A|\alpha\rangle$, then
\begin{align}
&\langle\chi_{l\alpha}|O^\alpha_{0,\mathrm{I}}|\chi_{m\alpha}\rangle t_{m\alpha}\nonumber\\
&=f_L^D(\alpha)\sum_{B>C, J>K}\langle\alpha|a^I_A a^L_D a^{BC}_{JK}|\alpha\rangle t^{JK}_{BC}(\alpha)\nonumber\\
&=f_J^B(\alpha)t^{IJ}_{AB}(\alpha).\label{SD}
\end{align}
On the other hand, if $|\chi_{l\alpha}\rangle=a_{IJ}^{AB}|\alpha\rangle$ with $A>B$ and $I>J$, then
\begin{align}
&\langle\chi_{l\alpha}|O^\alpha_{0,\mathrm{I}}|\chi_{m\alpha}\rangle t_{m\alpha}\nonumber\\
&=f_D^L(\alpha)\langle\alpha|a^{IJ}_{AB}a^D_L a_K^C|\alpha\rangle t^{K}_{C}(\alpha)\nonumber\\
&=f_A^I(\alpha)t^{J}_{B}(\alpha)+f_B^J(\alpha)t_A^J(\alpha)\nonumber\\
&=\mathcal{S}^{AB}_{IJ}f_A^I(\alpha)t^{J}_{B}(\alpha),\label{DS}
\end{align}
where the symmetrizer reads $\mathcal{S}^{AB}_{IJ}=1+(AB)(IJ)$.
Eqs. \eqref{SD} and \eqref{DS} describe the couplings between external singles and doubles (SD) and
between external doubles and singles (DS), respectively.

The collection of Eqs. \eqref{SS}, \eqref{DD}, \eqref{SD} and \eqref{DS}
completes the expression for the matrix elements of $[\tilde{H}_0^\alpha,T^\alpha]$ with the following
all-class zeroth-order Hamiltonian $\tilde{H}_0^\alpha$
\begin{align}
\tilde{H}_0^\alpha&= \tilde{H}_{0,\mathrm{I}}^\alpha + H_{0,\mathrm{A}}^\alpha,\label{All-classH0}\\
\tilde{H}_{0,\mathrm{I}}^\alpha&= f_I^I(\alpha) + Qf^\alpha Q,\label{All-classH0I}
\end{align}
where $Qf^\alpha Q$ is in place of the last two terms of $H_{0,\mathrm{I}}^\alpha$ in Eq. \eqref{H0Inactive}.
While such replacement does introduce additional couplings between external functions that are missed by the class-diagonal
operator $H_{0,\mathrm{I}}^\alpha$ \eqref{H0Inactive},
the DS type of couplings arising from single excitations of singles
is manifestly disconnected [cf. Eq. \eqref{DS}], so that the size-extensivity is lost.
However, the all-class $\tilde{H}_0^\alpha$ \eqref{All-classH0} is advantageous in that
spin adaptation (which requires all types of couplings) can be achieved
even when individual NEDs are taken as the basis. Moreover, the formally violated size-consistency
can also be resumed in practice (see Sec. \ref{SecSizeCon}).
The same occur also to the `standard' all-class zeroth-order Hamiltonian defined in Eq. \eqref{ReferenceH0}.

\subsubsection{Size-consistency}\label{SecSizeCon}
Different from size-extensivity\cite{Size-extensivity} that applies to theoretical methods
without reference to any type of physical systems, size-consistency\cite{Size-consistency} is
defined as the additive separability of the energy for a special type of physical systems:
molecular systems AB composed of two physically separated, non-interacting subsystems A and B.
Although the connectedness of the $\mathbf{t}$-amplitudes,
combined with the use of a model space $\mathcal{M}_0(\mathrm{AB})$ for the supermolecule AB
that is the direct product [denoted as $\mathcal{M}_0(\mathrm{A})\otimes \mathcal{M}_0(\mathrm{B})$] of
those of the non-interacting subsystems A and B, already implies that PASPT2 is size-consistent,
we decide to demonstrate this explicitly, without explicit reference to the connectivity of the $\mathbf{t}$-amplitudes.
To that end, assume first that the first-order $T^{\alpha_{\mathrm{AB}}}$-cluster operator of AB takes the following form
\begin{align}
T^{\alpha_{\mathrm{AB}}}&=T^{\alpha_{\mathrm{A}}}\otimes 1_{\mathrm{B}}+ 1_{\mathrm{A}}\otimes T^{\alpha_{\mathrm{B}}}+
\bar{T}^{\alpha_{\mathrm{AB}}},\label{amplitude_AB}\\
T^{\alpha_{\mathrm{Y}}}&=\sum_{X_{\mathrm{Y}}} t_{X_{\mathrm{Y}}}^{\alpha_{\mathrm{Y}}}\{X^\dag_{\mathrm{Y}}\}_{\alpha_{\mathrm{Y}}}, \quad\mbox{ Y = A, B},\\
\bar{T}^{\alpha_{\mathrm{AB}}}&=\sum_{X_{\mathrm{A}}X_{\mathrm{B}}}t_{X_{\mathrm{A}}X_{\mathrm{B}}}^{\alpha_{\mathrm{A}}\alpha_{\mathrm{B}}}
\{X_{\mathrm{A}}^\dag\otimes X_{\mathrm{B}}^\dag\}_{\alpha_{\mathrm{A}}\alpha_{\mathrm{B}}},
\end{align}
where $X^\dag_{\mathrm{A}}$ and $X^\dag_{\mathrm{B}}$
represent the single and double excitations on A and B, respectively, whereas $X_{\mathrm{A}}^\dag\otimes X_{\mathrm{B}}^\dag$ represents
their direct products.
Since the operators $O_{\mathrm{AB}}$ of the supermolecule are additively separable
\begin{align}
H_{\mathrm{AB}}&=H_{\mathrm{A}}\otimes 1_{\mathrm{B}}+ 1_{\mathrm{A}}\otimes H_{\mathrm{B}},\label{HAB}\\
H_0^{\alpha_{\mathrm{AB}}}&=H_0^{\alpha_{\mathrm{A}}}\otimes 1_{\mathrm{B}}+ 1_{\mathrm{A}}\otimes H_0^{\alpha_{\mathrm{B}}},\label{H0AB} \\
V^{\alpha_{\mathrm{AB}}}&=V^{\alpha_{\mathrm{A}}}\otimes 1_{\mathrm{B}}+ 1_{\mathrm{A}}\otimes V^{\alpha_{\mathrm{B}}},\label{VAB}
\end{align}
we have
\begin{align}
\langle\alpha_{\mathrm{A}}\alpha_{\mathrm{B}}|X_{\mathrm{A}}\otimes X_{\mathrm{B}} O_{\mathrm{AB}}|\alpha_{\mathrm{A}}\alpha_{\mathrm{B}}\rangle\equiv 0 \label{ZeroOAB}
\end{align}
for all collective excitations on A and B, whether external or internal.
It follows that the collective-excitation amplitudes $t_{X_{\mathrm{A}}X_{\mathrm{B}}}^{\alpha_{\mathrm{A}}\alpha_{\mathrm{B}}}$
are not coupled to the on-site ones $t_{X_{\mathrm{Y}}}^{\alpha_{\mathrm{Y}}}$ (Y = A, B) and are also source free
[i.e., $V_{X_{\mathrm{A}}X_{\mathrm{B}}}^{\alpha_{\mathrm{A}}\alpha_{\mathrm{B}}}=\langle\alpha_{\mathrm{A}}\alpha_{\mathrm{B}}|X_{\mathrm{A}}\otimes X_{\mathrm{B}} V^{\alpha_{\mathrm{A}}\alpha_{\mathrm{B}}}|\alpha_{\mathrm{A}}\alpha_{\mathrm{B}}\rangle=0$].
That is, the linear system of equations (of the form $\mathbf{A}\mathbf{t}=\mathbf{0}$) for
$t_{X_{\mathrm{A}}X_{\mathrm{B}}}^{\alpha_{\mathrm{A}}\alpha_{\mathrm{B}}}$ are \emph{homogeneous},
so that $t_{X_{\mathrm{A}}X_{\mathrm{B}}}^{\alpha_{\mathrm{A}}\alpha_{\mathrm{B}}}$ must be zero.
Therefore, $T^{\alpha_{\mathrm{AB}}}$ can be simplified to
\begin{align}
T^{\alpha_{\mathrm{AB}}}&=T^{\alpha_{\mathrm{A}}}\otimes 1_{\mathrm{B}}+ 1_{\mathrm{A}}\otimes T^{\alpha_{\mathrm{B}}}.\label{amplitude_ABu}
\end{align}
The commutator $[H_0^{\alpha_{\mathrm{AB}}},T^{\alpha_{\mathrm{AB}}}]$ is then also additively separable
\begin{align}
[H_0^{\alpha_{\mathrm{AB}}},T^{\alpha_{\mathrm{AB}}}]&=[H_0^{\alpha_{\mathrm{A}}},T^{\alpha_{\mathrm{A}}}]\otimes 1_{\mathrm{B}}+ 1_{\mathrm{A}}\otimes
[H_0^{\alpha_{\mathrm{B}}},T^{\alpha_{\mathrm{B}}}],\label{H0Tab}
\end{align}
and meanwhile satisfies the identity in Eq. \eqref{ZeroOAB}.

For on-site internal excitations $X_{\mathrm{Y}}^\dag$ (Y = A, B), the subsystem C-conditions
\begin{align}
\langle\alpha_{\mathrm{Y}}|X_{\mathrm{Y}}T^{\alpha_{\mathrm{Y}}}|\alpha_{\mathrm{Y}}\rangle=0
\end{align}
implies that the C-conditions for AB are also satisfied, i.e.,
\begin{align}
\langle\alpha_{\mathrm{A}}\alpha_{\mathrm{B}}|X_{\mathrm{Y}}T^{\alpha_{\mathrm{AB}}}|\alpha_{\mathrm{A}}\alpha_{\mathrm{B}}\rangle
=\langle\alpha_{\mathrm{Y}}|X_{\mathrm{Y}}T^{\alpha_{\mathrm{Y}}}|\alpha_{\mathrm{Y}}\rangle=0.
\end{align}
Likewise, for on-site external excitations $X_{\mathrm{Y}}^\dag$ (Y = A, B), we have
\begin{align}
\langle\alpha_{\mathrm{A}}\alpha_{\mathrm{B}}|X_{\mathrm{Y}}O_{\mathrm{AB}}|\alpha_{\mathrm{A}}\alpha_{\mathrm{B}}\rangle
=\langle\alpha_{\mathrm{Y}}|X_{\mathrm{Y}}O_{\mathrm{Y}}|\alpha_{\mathrm{Y}}\rangle
\end{align}
for any operator in Eqs. \eqref{HAB}, \eqref{H0AB}, \eqref{VAB} and \eqref{H0Tab}. It follows that
inserting $T^{\alpha_{\mathrm{AB}}}$ \eqref{amplitude_ABu} into
Eq. \eqref{1stEq} for the amplitudes of AB gives rises to
two uncoupled sets of amplitude equations, one for A and the other for B.
Therefore, given the $\mathbf{t}$-amplitudes of A (denoted as $\mathbf{t}_{\mathrm{A}}$)
and B (denoted as $\mathbf{t}_{\mathrm{B}}$) by separate calculations, those of AB  can be obtained simply as
$\mathbf{t}_{\mathrm{A}}\otimes\mathbf{I}_{\mathrm{B}}+\mathbf{I}_{\mathrm{A}}\otimes\mathbf{t}_{\mathrm{B}}$.

The effective Hamiltonian matrix elements for AB read
\begin{align}
H^{\mathrm{eff,AB}[2]}_{\beta_{\mathrm{A}}\beta_{\mathrm{B}},\alpha_{\mathrm{A}}\alpha_{\mathrm{B}}}
&=\langle\beta_{\mathrm{A}}\beta_{\mathrm{B}}|H_{\mathrm{AB}}|\alpha_{\mathrm{A}}\alpha_{\mathrm{B}}\rangle
+\langle\beta_{\mathrm{A}}\beta_{\mathrm{B}}
|H_{\mathrm{AB}} QT^{\alpha_{\mathrm{A}}\alpha_{\mathrm{B}}}|\alpha_{\mathrm{A}}\alpha_{\mathrm{B}}\rangle\nonumber\\
&=\langle\beta_{\mathrm{A}}|H_{\mathrm{A}}|\alpha_{\mathrm{A}}\rangle\delta_{\beta_{\mathrm{B}},\alpha_{\mathrm{B}}}+
\langle\beta_{\mathrm{B}}|H_{\mathrm{B}}|\alpha_{\mathrm{B}}\rangle\delta_{\beta_{\mathrm{A}},\alpha_{\mathrm{A}}}\nonumber\\
&+\sum_{X_\mathrm{A}}\langle\beta_{\mathrm{A}}\beta_{\mathrm{B}}
|H_{\mathrm{AB}} Q X_{\mathrm{A}}^\dag|\alpha_{\mathrm{A}}\alpha_{\mathrm{B}}\rangle
\langle\alpha_{\mathrm{A}}\alpha_{\mathrm{B}}|X_{\mathrm{A}}T^{\alpha_{\mathrm{A}}\alpha_{\mathrm{B}}}|\alpha_{\mathrm{A}}\alpha_{\mathrm{B}}\rangle \nonumber\\
&+\sum_{X_\mathrm{B}}\langle\beta_{\mathrm{A}}\beta_{\mathrm{B}}
|H_{\mathrm{AB}} Q X_{\mathrm{B}}^\dag|\alpha_{\mathrm{A}}\alpha_{\mathrm{B}}\rangle
\langle\alpha_{\mathrm{A}}\alpha_{\mathrm{B}}|X_{\mathrm{B}}T^{\alpha_{\mathrm{A}}\alpha_{\mathrm{B}}}|\alpha_{\mathrm{A}}\alpha_{\mathrm{B}}\rangle\nonumber\\
&=\delta_{\beta_{\mathrm{B}},\alpha_{\mathrm{B}}}\langle\beta_{\mathrm{A}}|H_{\mathrm{A}}|\alpha_{\mathrm{A}}\rangle+
\delta_{\beta_{\mathrm{A}},\alpha_{\mathrm{A}}}\langle\beta_{\mathrm{B}}|H_{\mathrm{B}}|\alpha_{\mathrm{B}}\rangle\nonumber\\
&+\delta_{\beta_{\mathrm{B}},\alpha_{\mathrm{B}}}\sum_{X_\mathrm{A}}\langle\beta_{\mathrm{A}}
|H_{\mathrm{A}} Q X_{\mathrm{A}}^\dag|\alpha_{\mathrm{A}}\rangle
\langle\alpha_{\mathrm{A}}|X_{\mathrm{A}}T^{\alpha_{\mathrm{A}}}|\alpha_{\mathrm{A}}\rangle\nonumber\\
&+\delta_{\beta_{\mathrm{A}},\alpha_{\mathrm{A}}}\sum_{X_\mathrm{B}}\langle\beta_{\mathrm{B}}
|H_{\mathrm{B}} Q X_{\mathrm{B}}^\dag|\alpha_{\mathrm{B}}\rangle
\langle\alpha_{\mathrm{B}}|X_{\mathrm{B}}T^{\alpha_{\mathrm{B}}}|\alpha_{\mathrm{B}}\rangle\nonumber\\
&=\delta_{\beta_{\mathrm{B}},\alpha_{\mathrm{B}}}H^{\mathrm{eff,A}[2]}_{\beta_{\mathrm{A}},\alpha_{\mathrm{A}}}+
\delta_{\beta_{\mathrm{A}},\alpha_{\mathrm{A}}}H^{\mathrm{eff,B}[2]}_{\beta_{\mathrm{B}},\alpha_{\mathrm{B}}},
\end{align}
which shows clearly that $H^{\mathrm{eff,AB}[2]}$ \eqref{PASH2a} is additively separable, even though
$H^{\mathrm{eff,A}[2]}$ and/or $H^{\mathrm{eff,B}[2]}$
may contain disconnected terms. It then follows trivially that
the energies $\{E_{AB}(ij)\}_{(ij)=1}^{M_AM_B}$ calculated
for the whole AB system
are precisely $\{E_{\mathrm{A}}(i)+E_{\mathrm{B}}(j)\}_{(ij)=1}^{M_AM_B}$,
with $\{ E_{\mathrm{A}}(i)\}_{i=1}^{M_{\mathrm{A}}}$ and
$\{E_{\mathrm{B}}(j)\}_{j=1}^{M_{\mathrm{B}}}$ being the respective energies of A and B calculated separately.

What is essential for the above proof of the size-consistency of PASPT2 lies in the additive separability
of the zeroth-order Hamiltonian $H_0^\alpha$ \eqref{H0final}, instead of the connectivity of the $\mathbf{t}$-amplitudes and the effective Hamiltonian.
The situation would be different if a nonseparable $H_0^\alpha$ is to be adopted, e.g., the projected one\cite{QFQ1987}
\begin{align}
H^\alpha_0&=PF^\alpha P+QF^\alpha Q, \quad F^\alpha=F_p^q(\alpha) a^p_q, \label{ReferenceH0}
\end{align}
where $F^\alpha$ may be, e.g., a spin-averaged Fock operator. $H_0^{\alpha_{\mathrm{AB}}}$ \eqref{ReferenceH0} can be decomposed as
\begin{align}
H_0^{\alpha_{\mathrm{AB}}}
&=H_0^{\alpha_{\mathrm{A}}}\otimes 1_{\mathrm{B}}+ 1_{\mathrm{A}}\otimes H_0^{\alpha_{\mathrm{B}}}+ \bar{H}_0^{\alpha_{\mathrm{AB}}}, \\
H_0^{\alpha_{\mathrm{Y}}}&=P_{\mathrm{Y}}F^{\alpha_{\mathrm{Y}}}P_{\mathrm{Y}}+Q_{\mathrm{Y}}F^{\alpha_{\mathrm{Y}}}Q_{\mathrm{Y}},\quad \mathrm{Y = A, B},\\
\bar{H}_0^{\alpha_{\mathrm{AB}}}&=(P_{\mathrm{A}}F^{\alpha_{\mathrm{A}}} Q_{\mathrm{A}}+\mathrm{h.c.})\otimes Q_{\mathrm{B}}+
Q_{\mathrm{A}}\otimes (P_{\mathrm{B}}F^{\alpha_{\mathrm{B}}} Q_{\mathrm{B}}+\mathrm{h.c.})\label{NonsepH0}
\end{align}
by making use of the following identities
\begin{align}
Q_{\mathrm{AB}}&=1_{\mathrm{A}}\otimes 1_{\mathrm{B}}-P_{\mathrm{A}}\otimes P_{\mathrm{B}}\nonumber\\
&=Q_{\mathrm{A}}\otimes 1_{\mathrm{B}}+P_{\mathrm{A}}\otimes Q_{\mathrm{B}}\nonumber\\
&=1_{\mathrm{A}}\otimes Q_{\mathrm{B}}+Q_{\mathrm{A}}\otimes P_{\mathrm{B}}.\label{QABDeompos}
\end{align}
The first-order interaction operator then reads
\begin{align}
V^{\alpha_{\mathrm{AB}}}
&=V^{\alpha_{\mathrm{A}}}\otimes 1_{\mathrm{B}}+ 1_{\mathrm{A}}\otimes V^{\alpha_{\mathrm{B}}} - \bar{H}_0^{\alpha_{\mathrm{AB}}}\label{NosepVAB}.
\end{align}
Straightforward calculations give rise to the following system of linear equations for the amplitudes of AB
\begin{subequations}\label{CollectiveEq}
\begin{align}
&\sum_{Y_{\mathrm{A}}}A^{\alpha_{\mathrm{A}}\alpha_{\mathrm{A}}}_{X_{\mathrm{A}}Y_{\mathrm{A}}}
t^{\alpha_{\mathrm{A}}}_{Y_{\mathrm{A}}}-\sum_{Y_{\mathrm{B}}}
t^{\alpha_{\mathrm{A}}\alpha_{\mathrm{B}}}_{X_{\mathrm{A}}Y_{\mathrm{B}}}(S^{\alpha_{\mathrm{B}}}_{Y_{\mathrm{B}}})^\dag
=V_{X_{\mathrm{A}}}^{\alpha_{\mathrm{A}}},\label{Aeq}\\
&\sum_{Y_{\mathrm{B}}}A^{\alpha_{\mathrm{B}}\alpha_{\mathrm{B}}}_{X_{\mathrm{B}}Y_{\mathrm{B}}}
t^{\alpha_{\mathrm{B}}}_{Y_{\mathrm{B}}}-\sum_{Y_{\mathrm{A}}}
t^{\alpha_{\mathrm{B}}\alpha_{\mathrm{A}}}_{X_{\mathrm{B}}Y_{\mathrm{A}}}(S^{\alpha_{\mathrm{A}}}_{Y_{\mathrm{A}}})^\dag
=V_{X_{\mathrm{B}}}^{\alpha_{\mathrm{B}}},\label{Beq}\\
&\sum_{Y_{\mathrm{A}}}A^{\alpha_{\mathrm{A}}\alpha_{\mathrm{A}}}_{X_{\mathrm{A}}Y_{\mathrm{A}}}t^{\alpha_{\mathrm{B}}\alpha_{\mathrm{A}}}_{X_{\mathrm{B}}Y_{\mathrm{A}}}
+\sum_{Y_{\mathrm{B}}}t^{\alpha_{\mathrm{A}}\alpha_{\mathrm{B}}}_{X_{\mathrm{A}}Y_{\mathrm{B}}}A^{\alpha_{\mathrm{B}}\alpha_{\mathrm{B}}}_{X_{\mathrm{B}}Y_{\mathrm{B}}}
-S_{X_{\mathrm{A}}}^{\alpha_{\mathrm{A}}}t_{X_{\mathrm{B}}}^{\alpha_{\mathrm{B}}}
-t_{X_{\mathrm{A}}}^{\alpha_{\mathrm{A}}}S_{X_{\mathrm{B}}}^{\alpha_{\mathrm{B}}}=V^{\alpha_{\mathrm{A}}\alpha_{\mathrm{B}}}_{X_{\mathrm{A}}X_{\mathrm{B}}}=0,
\end{align}
\end{subequations}
where
\begin{align}
V_{X_{\mathrm{Y}}}^{\alpha_{\mathrm{Y}}}&=\langle\alpha_{\mathrm{Y}}|X_{\mathrm{Y}}V^{\alpha_{\mathrm{Y}}} |\alpha_{\mathrm{Y}}\rangle,\quad \mbox{ Y = A, B},\\
S^{\alpha_{\mathrm{Y}}}_{X_{\mathrm{Y}}}&=\langle\alpha_{\mathrm{Y}}| X_{\mathrm{Y}}F_{\mathrm{Y}}^{\alpha_{\mathrm{Y}}}|\alpha_{\mathrm{Y}}\rangle,\\
V^{\alpha_{\mathrm{A}}\alpha_{\mathrm{B}}}_{X_{\mathrm{A}}X_{\mathrm{B}}}&=\langle\alpha_{\mathrm{A}}\alpha_{\mathrm{B}}|
X_{\mathrm{A}}\otimes X_{\mathrm{B}}H_{AB}|\alpha_{\mathrm{A}}\alpha_{\mathrm{B}}\rangle.
\end{align}
The $\mathbf{A}$ matrices in Eq. \eqref{CollectiveEq} for the subsystems  take the same form as that in Eq. \eqref{LinearSys}.
It is seen that the (still source-free) collective-excitation amplitudes $\mathbf{t}_{X_{\mathrm{A}}X_{\mathrm{B}}}^{\alpha_{\mathrm{A}}\alpha_{\mathrm{B}}}$
(for singles on A and singles on B)
are nonvanishing and coupled to the on-site ones $\mathbf{t}_{X_{\mathrm{Y}}}^{\alpha_{\mathrm{Y}}}$ (Y = A, B). Consequently,
the latter are no longer identical to those obtained by separate calculations of subsystems A and B, which renders
the energies of AB different from the sums of those of subsystems A and B (NB:
the collective-excitation amplitudes $\mathbf{t}_{X_{\mathrm{A}}X_{\mathrm{B}}}^{\alpha_{\mathrm{A}}\alpha_{\mathrm{B}}}$ do not contribute to the energies of AB).
Nevertheless, it can be noted that the \emph{source-free} collective excitations
can readily be screened out in practice (see Sec. \ref{Implementation}), so as to resume the size-consistency of the calculated energies (vide post).
However, the projected $H_0^\alpha$ \eqref{ReferenceH0} does destroy the size-extensivity of PASPT2 (see Sec. \ref{SecCommutator}).

\subsubsection{Comparison with SS-MRPT2 and JM-MRPT2/JM-HeffPT2}
A brief comparison of PASPT2 with SS-MRPT2\cite{SS-MRPT2a,SS-MRPT2b} and JM-MRPT2/JM-HeffPT2\cite{JM-MRPT2} should be made.
They are common in the following aspects: (a) they are all rooted in the multi-exponential JM Ansatz\cite{JM1981} for the wave operator;
(b) they all take individual electron configurations (determinants or configuration state functions) as perturbers;
(c) they all can capture the full relaxation of the reference coefficients
under the influence of dynamic correlation, with the exception of JM-MRPT2; (d) they are all size-extensive and size-consistent;
(e) they are all non-invariant with respect to orbital rotations,
due to dependence of the approximate $\mathbf{t}$-amplitudes on the orbital non-variance of the reference functions.
However, they are also distinct in several aspects:
(1) both SS-MRPT2 and JM-MRPT2/JM-HeffPT2 are based on a CAS and state specific, whereas PASPT2 is based on a PAS/IAS and state universal;
(2) both SS-MRPT2 and JM-MRPT2/JM-HeffPT2 employ state-specific CASSCF energies to avoid too small energy denominators and hence the intruder-state problem
for low-lying states, whereas PASPT2 achieves this even for high-lying states by introducing a buffer that is determined automatically
for a given PAS to form a closed intermediate Hamiltonian $\bar{H}^{\mathrm{eff}[2]}_X$
(NB: in case that $\bar{\mathcal{M}}_0=\emptyset$ (for $\bar{H}^{\mathrm{eff}[2]}_Q$), those
external functions with small energy denominators can simply be removed);
(3) both SS-MRPT2 and PASPT2 have to solve iteratively a system of linear equations for the $\mathbf{t}$-amplitudes,
whereas JM-MRPT2/JM-HeffPT2 employs a reference-specific diagonal zeroth-order Hamiltonian, so that the $\mathbf{t}$-amplitudes
can be calculated directly;
(4) The amplitude equations of PASPT2 and SS-MRPT2 are very similar in structure. However,
the third, coupling term in Eq. \eqref{MP1}, resulting from the non-diagonality of $H_0^\alpha$ within the $P$-space,
is fundamentally different from the counterpart in SS-MRPT2 [cf. the last term in equation (11) of Ref. \citenum{MaoSS-MRPT2a}];
(5) While SS-MRPT2 and JM-HeffMRPT2 belong to the static-dynamic-static family of methods,
JM-MRPT2 and PASPT2 belong to the static-then-dynamic and dynamic-then-static families of methods, respectively\cite{SDS,iCI}.

\section{Implementation}\label{Implementation}

Since the NEDs generated from the same orbital configuration (oCFG) have the same zeroth-order energy,
they should not be split into the $P$- and $Q$-spaces, to avoid zero energy denominators. Therefore, the PAS $\mathcal{M}_0$ should consist
of all NEDs with a specified spin projection $M_s$ from the reference oCFGs. Likewise, the $Q$-space
can be constructed by generating all NEDs with the same $M_s$ from the singly and doubly excited oCFGs.
In essence, it is oCFGs that are taken here as the organization units.
Note in passing that some external NEDs generated this way may appear as higher-than-double excitations from the reference NEDs
(which are necessary for spin adaptation of open-shell electron configurations), for an excited oCFG
may have two or four more open-shell orbitals, some of which can have spins opposite to those of the same open-shell orbitals in the parent reference oCFG,
thereby requiring additional same-spin excitations
(NB: the assignment of electron spin to open-shell orbitals is constrained only by the $M_s$ value).
Still, however, spin adaptation cannot be achieved, unless
the class-diagonal inactive operator $H_{0,\mathrm{I}}^\alpha$ \eqref{H0Inactive} is replaced with
the all-class inactive operator $\tilde{H}_{0,\mathrm{I}}^\alpha$ \eqref{All-classH0I}.
In the latter case, all states with spins $S\in[M_s, \frac{M_o}{2}]$ can be captured. Here,
$M_o$ is the maximum number of open-shell orbitals in the reference oCFGs.
However, this comes with a cost to size-extensivity (see Sec. \ref{SecCommutator}).
Therefore, the explicit spin adaptation of PASPT2 is mandatory but which goes beyond the present scope.

The corresponding matrix elements of $H_0^\alpha$ and $V^\alpha$
can readily be evaluated with the \texttt{Hamiltonian} module in \texttt{MetaWave}\cite{MetaWave},
a C++ template-based platform for rapid development and high-performance execution of wave function-based quantum chemical methods.
Eq. \eqref{LinearSys} can then be solved iteratively with the generalized minimal residual method\cite{GMRES} as follows:
\begin{enumerate}[(1)]
\item Initial guess: calculate the MP1 amplitudes $t^0_{l\alpha}=-V_{l\alpha}/\Delta E_{l\alpha}$ for each reference $|\alpha\rangle$.
If $|t^0_{l\alpha}|$ is less than $\tau_S$ (e.g., $10^{-8}$) or larger than $\tau_L$ (e.g., $0.3$),
discard the corresponding external function $|\chi_{l\alpha}\rangle$. The latter is to remove
nonphysical contributions due to the violation of the Brillouin conditions.
\item LCUT: expand the $\mathbf{t}$-amplitudes as a linear combination of the uncoupled $\mathbf{t}^0$-amplitudes (i.e.,
$\mathbf{t}=\sum_\alpha\mathbf{t}^0_\alpha \mathbf{c}_\alpha$) and determine the expansion coefficients
by minimizing the cost function $\|\mathbf{A}\mathbf{t}- \mathbf{V}\|_2$. For most purposes,
the LCUT approximation is already sufficient.
\item Iteration: starting with the LCUT, both the residual $\mathbf{r}_i=\mathbf{A}\mathbf{t}_i - \mathbf{V}$ and the
preconditioned residual $-(\boldsymbol{\Delta E})^{-1}\mathbf{r}_i$ are added to the Krylov subspace of the previous iterations,
after orthonormalizing them against the available vectors.
The iteration is restarted every $n$ steps, so that the Krylov subspace contains at most $2n$ vectors, with $n$ being around 15.
The convergence is reached if $||\mathbf{A}\mathbf{t} - \mathbf{V}||_2 < \eta =10^{-9}$.
\end{enumerate}
At this moment, two points should be mentioned. Firstly,
the two-body term in $H_{0,\mathrm{A}}^{\alpha}$ \eqref{H0Active}
plays only a minor role and can hence be estimated by performing just one iteration upon convergence of Eq. \eqref{LinearSys} only with the one-body term in $H_{0,\mathrm{A}}^{\alpha}$ \eqref{H0Active}. Secondly, the convergence may be slowed down significantly by the existence of
anti-correlating components (i.e., those external functions with positive energy denominators). Since
the corresponding reference NEDs contribute significantly only to some high-lying states but not to the lowest ones,
such anti-correlating components can be screened out from the outset if only the lowest states are wanted.
On the other hand, if the high-lying states are also of interest, their amplitudes should be sought for directly by
bypassing the low-lying ones, following, e.g., the iterative vector interaction approach\cite{iVI,iVI-TDDFT}.

Overall, the computation is dominated by the matrix-vector product $\mathbf{At}$, which
scales as $\mathcal{O}(N_{it} M K^5)$, roughly $N_{it}$ times $M$ parallel, non-Brillouin MP2 calculations.
Here, $N_{it}$ is the number of iterations required to converge Eq. \eqref{LinearSys}.
If the LCUT approximation is adopted, the factor of $N_{it}$ does not apply.

Unless other stated, the cc-pVDZ basis set\cite{Dunning1989} is to used in subsequent calculations with the BDF program package\cite{BDF1,BDF2020}.

\section{Results and discussion}\label{Pilot}
\subsection{Numerical check of size-extensivity and size-consistency}
\begin{figure}[H]
	\centering
	\includegraphics[width=0.5\textwidth]{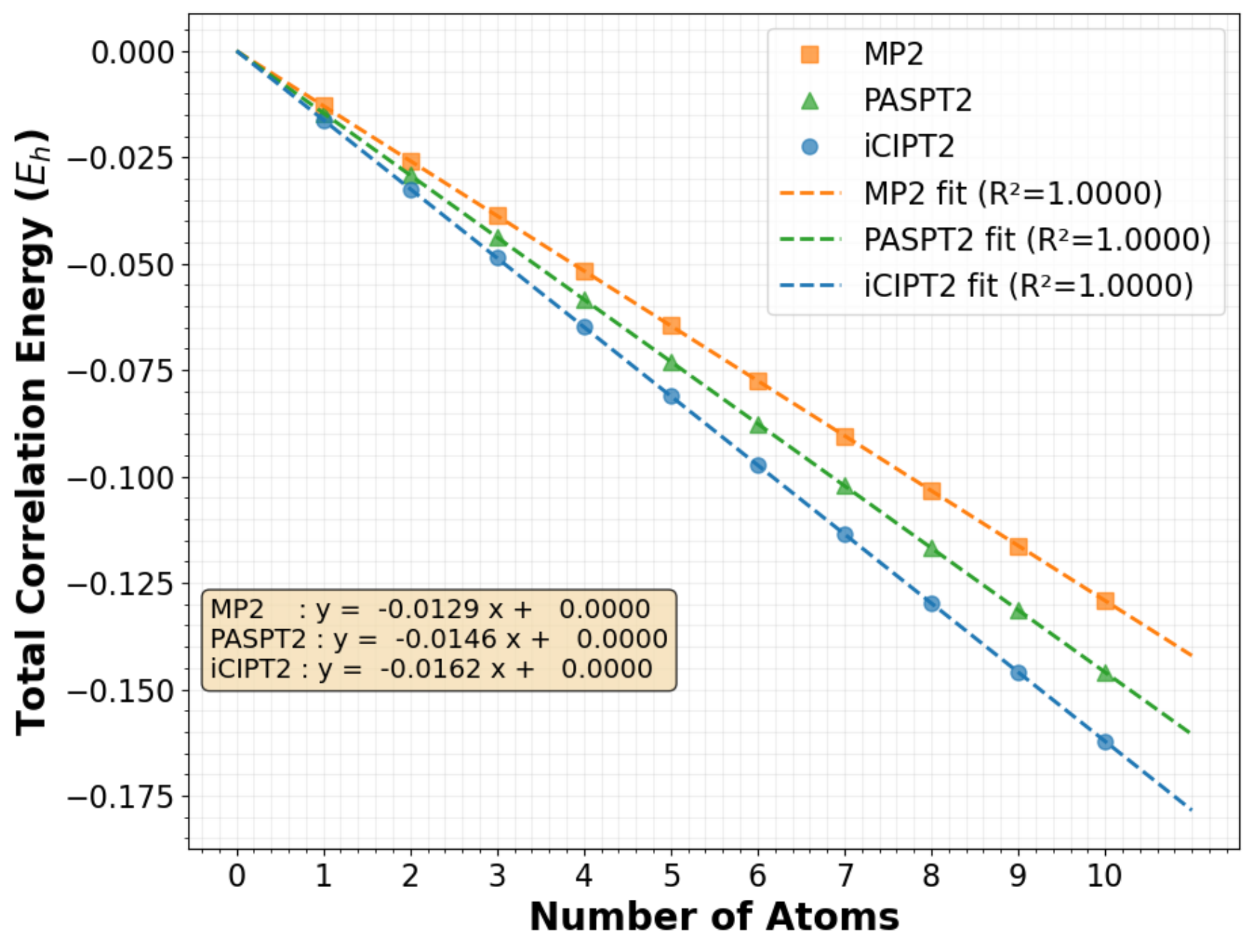}
	\caption{
		Correlation energies in linear \ce{He} chains by MP2, PASPT2 and iCIPT2.}
	\label{fig:size_extensive}
\end{figure}

A size-extensive method should yield a total correlation energy that grows linearly with the number of electrons
in a system composed of interacting identical subsystems. To check this for PASPT2,
linear He chains He$_n$ ($n\in[1,10]$) were taken as examples. The interatomic distance was set to  3~\AA.
For each He chain He$_m$, both the HF and PASPT2 calculations were done in the full basis of the 10 He atoms by including ghost functions for the atoms that were not
included explicitly. The occupied and virtual HF orbitals were first localized separately and assigned to individual He atoms.
The Fock operator was then projected onto the local virtual orbitals and diagonalized to obtain local canonical virtual orbitals for each He atom.
The $1s^2$ and $2s^2$ configurations were chosen for each He atom, so that
the model space $\mathcal{M}_0$ for He$_m$ is composed of $2^m$ tensor products.
It was verified that such model spaces are closed under the action of the effective Hamiltonian $\bar{H}^{\mathrm{eff}[2]}$ \eqref{PASH2}.
The PASPT2 correlation energy as function of the number of He atoms is plotted in Fig. \ref{fig:size_extensive}.
It is seen that the PASPT2 correlation energy is indeed linear with the number of
He atoms (or electrons), just like those of the MP2 and (near-exact) iCIPT2\cite{iCIPT2,iCIPT2New}.
In particular, the three lines share the same zero-electron limit, as should be.

To demonstrate the size-consistency of PASPT2 numerically,
we take a AB-system consisting of two non-interacting but different \ce{H2} molecules:
one with a bond length of 0.7 \AA~ (subsystem A) and the other with a bond length of 0.8 \AA~ (subsystem B). The four hydrogen atoms are placed on the x-axis,
with the coordinates being 0.0, 0.7, 10000.0, and 10000.8 \AA, respectively.
The following model spaces were considered for the subsystems:
\begin{itemize}
	\item $\mathcal{M}_{0,1} = \{|0\bar{0}\rangle,|1\bar{1}\rangle\}$;
	\item $\mathcal{M}_{0,2} = \{|0\bar{1}\rangle,|\bar{0}1\rangle,|1\bar{2}\rangle,|\bar{1}2\rangle\}$;
	\item $\mathcal{M}_{0,3} = \{|01\rangle,|12\rangle\}$;
	\item $\mathcal{M}_{0,4} = \{|\bar{0}\bar{1}\rangle,|\bar{1}\bar{2}\rangle\}$.
\end{itemize}
Here, the numbers 0, 1 and 2 denote the $1\sigma_g$, $1\sigma_u$ and $2\sigma_g$ molecular orbitals, respectively.
For states of zero spin projection, the following 44-dimensional model space for the supermolecue AB is sufficient
\begin{equation*}
	\small
	\begin{split}
		\mathcal{M}_0(\mathrm{AB}) & = [\mathcal{M}_{0,1}(\mathrm{A})\otimes\mathcal{M}_{0,1}(\mathrm{B})]
\bigcup[\mathcal{M}_{0,2}(\mathrm{A})\otimes\mathcal{M}_{0,2}(\mathrm{B})]
\bigcup[\mathcal{M}_{0,1}(\mathrm{A})\otimes\mathcal{M}_{0,2}(\mathrm{B})] \\
&\bigcup[\mathcal{M}_{0,2}(\mathrm{A})\otimes\mathcal{M}_{0,1}(\mathrm{B})]
\bigcup[\mathcal{M}_{0,3}(\mathrm{A})\otimes\mathcal{M}_{0,4}(\mathrm{A})]
\bigcup[\mathcal{M}_{0,4}(\mathrm{A})\otimes\mathcal{M}_{0,3}(\mathrm{A})].
	\end{split}
\end{equation*}
The differences between $E_{\mathrm{AB}}$ and $E_\mathrm{A}+E_\mathrm{B}$ for the 44 states are zero within numerical accuracy.
The same was also found when the projected $H_0^\alpha$ in Eq. \eqref{ReferenceH0} instead of that in Eq. \eqref{H0final} was used. In this case, however,
the (source-free) simultaneous single excitations on A and B have to be screened out explicitly
(which can readily be done; see Sec. \ref{Implementation}), as discussed in Sec. \ref{SecSizeCon}.
This exercise shows that numerically size-consistent results can sometimes be obtained even if the chosen zeroth-order
Hamiltonian (e.g., that in Eq. \eqref{ReferenceH0}) furnishes neither size-extensivity (see Sec. \ref{SecCommutator}) nor size-consistency (see Sec. \ref{SecSizeCon}).

\subsection{Vertical excitations of \ce{H2O}}

To compare directly with the vertical excitation energies (VEE) of \ce{H2O} calculated by
the parent IN-GMS-SU-CCSD\cite{GMS-SU-CCSDJCP2004}, we first took the same settings in the PASPT2 calculations.
At the experimental geometry (bond length 0.9575 \AA, bond angle 104.5$^\circ$),
the lowest 9 HF orbitals of \ce{H2O} form the sequence $1a_1$, $2a_1$, $1b_1$, $3a_1$, $1b_2$, $4a_1$, $2b_1$, $3b_1$ and $5a_1$,
with $1a_1$ kept frozen.
The manually selected\cite{GMS-SU-CCSDJCP2004}  model spaces $\mathcal{M}_0$/HF ($\mathcal{M}_0$ built up with ground state canonical HF orbitals)
consist of 10, 12, 8 and 14 NEDs for the $A_1$, $A_2$, $B_1$ and $B_2$ irreducible representations (irrep),
respectively, see Table \ref{table:ModelSpaceInfo}. They were extended automatically
to $\mathcal{M}_X$ to form a closed intermediate Hamiltonian $\bar{H}^{\mathrm{eff}[2]}_X$ \eqref{HeffX}.
The diagonalization of  $\bar{H}^{\mathrm{eff}[2]}_X$ gives rise to $|\mathcal{M}_X|$ states, but only
the $|\mathcal{M}_0|$ states that have the largest projections
on $\mathcal{M}_0$ are meaningful. In general, the $|\mathcal{M}_0|$ states are
lower in energy than all the rest for each irrep. If this is not the case,
it would imply that the chosen $\mathcal{M}_0$ is problematic, which
can be taken as another internal check of the PASPT2 calculation.
For \ce{H2O}, the chosen $\mathcal{M}_0$\cite{GMS-SU-CCSDJCP2004} supports in total 43 excited states. Their VEEs
are documented in Table \ref{table:h2o_results}, with the iCIPT2\cite{iCIPT2,iCIPT2New} results (with $C_{\text{min}}=10^{-6}$) taken as the benchmark.
For comparison, the CASSCF-SDSPT2\cite{SDSPT2} calculations were also performed.

It is first seen from Table \ref{table:h2o_results} that, although the use of the zeroth-order Hamiltonian \eqref{H0final}
cannot achieve spin adaption (see Sec. \ref{SecCommutator}), the expectation values of the $S^2$ operator over the left and right eigenvectors
of the intermediate Hamiltonian $\bar{H}^{\mathrm{eff}[2]}_X$ are very close to the exact values. Notwithstanding this,
PASPT2 is less accurate than IN-GMS-SU-CCSD and also CASSCF-SDSPT2 in terms of the mean absolute errors (MAE),
not only for the 31 high-lying states of VEEs higher than 18 eV
but also for the 12 low-lying states of VEEs below 18 eV.
There could be several reasons for this. First of all, the class-diagonal form of $H_0^\alpha$ \eqref{H0final}
certainly limits the accuracy. To check this, we performed calculations with the projected $H_0^\alpha$
defined in Eq. \eqref{ReferenceH0} but still with $f^\alpha$ defined in Eq. \eqref{Fspin}.
This is an all-class zeroth-order Hamiltonian (see Sec. \ref{SecCommutator}).
It is seen that the PASPT2 VEEs are on average improved significantly for all the 43 states, even
better than those by IN-GMS-SU-CCSD. However,
this comes with a cost to size-extensivity (see Sec. \ref{SecCommutator})
and is therefore not really commended. Secondly, the use of ground state HF orbitals for
all excited states may also be problematic. As can be seen from Table \ref{table:h2o_results},
the VEEs for the 12 low-lying states are indeed improved when semi-canonical SA-CASSCF(8,8) orbitals were used
(which were obtained by canonicalizing separately the doubly occupied, active and virtual orbitals of
CASSCF(8,8) calculations averaged equally over all the states of each irrep). Yet, the VEEs for
the high-lying states got worsened. 
Thirdly, there is no reason to believe that the manually selected model spaces\cite{GMS-SU-CCSDJCP2004} are appropriate for all the states.
To check this, we constructed enlarged model spaces $\mathcal{M}_0^\prime$ by retaining the determinants of CAS(8,8)
with coefficients larger than $10^{-2}$ in absolute value. It turns out that $\mathcal{M}_0^\prime$ does embody $\mathcal{M}_0$
as a subset for each irrep and the corresponding $\mathcal{M}_X$ is just $\mathcal{M}_C$ in this case, see Table \ref{table:ModelSpaceInfo}.
As expected, the results are now uniformly better and become
essentially the same as those by CASSCF-SDSPT2 and IN-GMS-SU-CCSD.

For a better understanding of the above trends, the VEEs of \ce{H2O} obtained by projecting the bare Hamiltonian onto the
various model spaces are shown in Table S1 and Fig. S1 in the Supporting Information. It can be seen that, taking
$\mathcal{M}_C$/CASSCF [CAS(8,8) built up with SA-CASSCF orbitals] as the reference,
$\mathcal{M}_0$/CASSCF reduces the MAE (1.5 eV) of $\mathcal{M}_0$/HF to the desired extent (0.2 eV) for the low-energy part, but still
has a sizable MAE (3.0 eV) for the high-energy part of the spectrum. In contrast,
$\mathcal{M}_0^\prime$/CASSCF further improves $\mathcal{M}_0$/CASSCF for the whole spectrum.
All these are in line with the behavior of PASPT2. The implication is that
any MRPT2 calculation should start with a high-quality model space.

Finally, the different flavors of PASPT2 (in conjunction with $H_0^\alpha$ \eqref{H0final}) are compared in Table \ref{table:MAXMAE}. It is seen that the
$\bar{H}^{\mathrm{eff}[2]}$ and $\bar{H}^{\mathrm{eff}[2]}_Q$ variants (in conjunction with $\mathcal{M}_0$/HF) also
perform fairly well for the low-lying states. However, they are hardly recommended because of their proneness
to intruder states (which are not present here though).
Extending the $\mathcal{M}_X$ of $\mathcal{M}_0$ to $\mathcal{M}_C$ does improve the results (compare the `a,g' and `a,d' rows in Table \ref{table:MAXMAE}),
but this may not be practical for overly complicated systems.
Therefore, from both theoretical and computational points of view, the $\bar{H}^{\mathrm{eff}[2]}_X$ variant
(in conjunction with selected $\mathcal{M}_0$ and automatically extended $\mathcal{M}_X$) is the right option.
As a final note, the LCUT approximation of the $\mathbf{t}$-amplitudes is sufficiently accurate (compare the `a,f' and `a,d' rows in Table \ref{table:MAXMAE})
and was hence employed in subsequent calculations of \ce{N2}.

\begin{table}
	\centering
	\caption{Model spaces $\mathcal{M}_0$ for \ce{H2O} with $M_S = 0$. }
	\begin{threeparttable}
		\begin{tabular}{ccccc}\toprule
	Irrep	&	$A_1$ & $A_2$
	& $B_1$
	& $B_2$ \\\toprule
	oCFG$^a$& $|\mathrm{HF}\rangle$       & $1b_2\rightarrow 2b_1$
	& $3a_1\rightarrow 2b_1$
	& $1b_2\rightarrow 4a_1$         \\
			& $3a_1\rightarrow 4a_1$      & $1b_2\rightarrow 3b_1$
			& $1b_1\rightarrow 4a_1$
			& $3a_11b_2\rightarrow 4a_1^2$   \\
			& $1b_1\rightarrow 2b_1$      & $3a_11b_2\rightarrow 4a_12b_1$
			& $3a_1\rightarrow 3b_1$
			& $1b_2\rightarrow 5a_1$         \\
			& $1b_2^2\rightarrow 4a_1^2$  & $1b_11b_2\rightarrow 4a_1^2$
			& $1b_2^2\rightarrow 4a_12b_1$
			& $1b_11b_2\rightarrow 4a_12b_1$ \\
			& $3a_1\rightarrow 5a_1$      &
			&
			& $3a_11b_2\rightarrow 2b_1^2$   \\
			& $3a_1^2\rightarrow 4a_1^2$  &
			&
			&                                \\
			& $1b_2^2\rightarrow 2b_1^2$  &
			&
			&                                \\ \hline
			$(|M_0|, |M_X|, |M_C|)^b$  & (10, 104, 1234) &  (12, 22, 1234)
			&  (8, 10, 1216)
			& (14, 104, 1216)                    \\
            $(|M_0^\prime|, |M_X^\prime|, |M_C|)^c$  & (257, 1234,1234) & (256,1234,1234)
            & (258, 1216,1216)
            & (232, 1216,1216)    \\ \hline
		\end{tabular}
	\begin{tablenotes}
            \item[a] Orbital configuration.
			\item[b] Sizes of $\mathcal{M}_0$, $\mathcal{M}_X$ (on which the intermediate Hamiltonian is closed), and CAS(8,8).
            \item[c] $\mathcal{M}^\prime_0$ is selected from CAS(8,8) with $C_{\mathrm{min}}=10^{-2}$.
		\end{tablenotes}
\end{threeparttable} \label{table:ModelSpaceInfo}
\end{table}

\begin{table}
	\tiny
	\centering
	\caption{Deviations ($\Delta E$ in eV) of IN-GMS-SU-CCSD, PASPT2 and CASSCF-SDSPT2 from iCIPT2 for the
		vertical excitation energies of low-lying states of \ce{H2O}.
	}
	\begin{threeparttable}
		\begin{tabular}{rrrrrrrrrrrc}\hline\hline
			\multirow{2}{*}{Irrep}
			& \multirow{2}{*}{iCIPT2}
			& \multicolumn{1}{c}{SU-CCSD}
			& \multicolumn{2}{c}{PASPT2\tnote{a}}
			& \multicolumn{2}{c}{PASPT2\tnote{a,b}}
			& \multicolumn{2}{c}{PASPT2\tnote{c}}
			& \multicolumn{2}{c}{PASPT2\tnote{c,d}}
			& SDSPT2
			\\ \cline{3-12}
			&
			& $\Delta E$
			& $\Delta E$ & $\langle S^2\rangle$\tnote{e}
			& $\Delta E$ & $\langle S^2\rangle$\tnote{e}
			& $\Delta E$ & $\langle S^2\rangle$\tnote{e}
			& $\Delta E$ & $\langle S^2\rangle$\tnote{e}
			& $\Delta E$ \\ \hline
			$^1A_1$ &10.868 &-0.501 & -0.468 & 0.00 &  0.055 & 0.00 & 0.133 & 0.00 &  0.106 & 0.00 & 0.392  \\
			&17.912 &0.025  & -0.270 & 0.00 &  0.066 & 0.00 & 0.179 & 0.00 &  0.209 & 0.00 & 0.280  \\
			&21.213 &-0.746 & -0.446 & 0.00 &  0.371 & 0.00 & 0.999 & 0.00 & -0.350 & 0.00 & 2.744 \\
			&25.749 &-1.159 & -0.215 & 0.00 & -0.268 & 0.00 & 0.610 & 0.00 & -0.620 & 0.00 & -1.455\\
			&26.228 &-1.076 & -0.280 & 0.00 &  0.172 & 0.00 & 1.486 & 0.00 & -0.502 & 0.01 & -1.221\\
			&27.282 &-0.428 & -0.687 & 0.00 &  0.371 & 0.00 & 3.539 & 0.00 & -0.917 & 0.01 & -0.376 \\\hline
			$^3A_1$ &9.996 &-0.571  & -0.441 & 2.00 &  0.083 & 2.00 & 0.176 & 2.00 &  0.200 & 1.99 & 0.440 \\
			&15.531&0.023   & 0.205  & 2.00 &  0.592 & 2.00 & 0.676 & 2.00 &  0.403 & 2.00 & 0.449 \\
			&24.293&-0.834  & 0.529  & 2.00 &  0.955 & 2.00 & 0.882 & 2.00 &  0.615 & 2.17 & 1.740 \\\hline
			
			$^1A_2$&10.276&-0.252 &-0.746  & 0.00 &  0.048 & 0.00  & 0.284  & 0.00 &  0.474 & 0.01 & 0.276 \\
			&21.613& 0.067 &-0.317  & 0.04 &  0.385 & 0.17  & 0.538  & 0.00 &  0.503 & 0.04 & 0.246 \\
			&22.836&-0.901 &-1.164  & 0.00 & -0.298 & 0.00  & 1.718  & 0.00 & -0.239 & 0.00 & 0.408 \\
			&25.684&-0.377 &-0.874  & 0.00 &  0.004 & 0.00  & 2.719  & 0.00 &  0.046 & 0.01 & 0.511 \\
			&27.787&-0.672 &-1.694  & 0.00 & -0.693 & 0.00  & 1.351  & 0.00 & -0.193 & 0.00 & 0.614 \\\hline
			$^3A_2$&9.866 &-0.243 &-0.742  & 2.00 &  0.053 & 2.00  & 0.214  & 2.00 &  0.463 & 1.99 & 0.254 \\
			&21.449&-0.336 &-0.785  & 1.99 &  0.141 & 1.97  & 0.452  & 2.00 &  0.345 & 2.00 & 0.293 \\
			&22.034&-0.416 &-0.776  & 1.94 & -0.142 & 1.78  & 1.735  & 2.00 & -0.071 & 1.96 & 0.389 \\
			&23.495&-0.725 &-1.093  & 1.99 & -0.129 & 1.97  & 2.008  & 2.00 & -0.091 & 2.00 & 0.488 \\
			&24.174&-0.591 &-1.106  & 1.99 & -0.236 & 1.96  & 2.156  & 2.00 & -0.105 & 1.99 & 0.504 \\
			&25.393&-0.729 &-1.589  & 2.00 & -0.620 & 1.99  & 1.716  & 2.00 & -0.205 & 2.00 & 0.781 \\\hline
			$^5A_2$&20.663&-0.961 &-1.350  & 6.00 & -0.395 & 6.00  & 1.626  & 6.00 & -0.208 & 6.00 & 0.453 \\\hline
			
			$^1B_1$&12.951&-0.511 &-0.489 & 0.00 &  0.193 & 0.00  & 0.261  & 0.00 &  0.494 & 0.00 & 0.418 \\
			&14.868& 0.011 &-0.190 & 0.00 &  0.232 & 0.00  & 0.315  & 0.00 &  0.506 & 0.00 & 0.430 \\
			&25.026&-0.045 &-0.699 & 0.00 &  0.336 & 0.00  & 0.754  & 0.00 &  0.625 & 0.00 & 0.512 \\
			&26.027&-0.423 &-0.836 & 0.00 & -0.283 & 0.00  & 1.367  & 0.00 &  2.259 & 0.01 & 0.988 \\\hline
			$^3B_1$&12.041&-0.509 &-0.480 & 2.00 &  0.188 & 2.00  & 0.184  & 2.00 &  0.516 & 2.00 & 0.424 \\
			&13.754&-0.088 &-0.047 & 2.00 &  0.391 & 2.00  & 0.455  & 2.00 &  0.523 & 2.00 & 0.467 \\
			&22.097&-0.661 &-1.259 & 2.00 & -0.284 & 2.00  & 1.476  & 2.00 & -0.378 & 2.00 & 0.293 \\
			&23.180&-0.082 &-0.118 & 2.00 &  0.450 & 2.00  & 0.967  & 2.00 &  2.930 & 2.00 & 0.498 \\\hline
			
			$^1B_2$&8.241 &-0.279 &-0.673 &0.01 & -0.002 & 0.00 & 0.225  & 0.00 &  0.168 & 0.01 & 0.218 \\
			&22.886&-0.942 &-0.732 &0.00 &  0.048 & 0.01 & 0.913  & 0.00 & -0.273 & 0.00 & 0.473 \\
			&23.411&-1.051 &-0.118 &0.01 &  0.424 & 0.75 & 0.783  & 0.00 &  2.822 & 0.02 & 0.149 \\
			&27.168&-0.387 &-0.659 &0.00 & -0.174 & 0.05 & 1.112  & 0.01 &  0.161 & 0.04 & 0.427 \\
			&28.779&-0.749 &-1.706 &0.00 & -1.368 & 0.03 & 2.575  & 0.00 & -0.045 & 0.37 & 0.541 \\
			&30.551&-0.254 &-0.833 &0.00 & -0.011 & 0.01 & 1.795  & 0.00 &  0.077 & 0.00 & 0.633 \\\hline
			$^3B_2$&7.572 &-0.272 &-0.683 &1.99 &  0.004 & 2.00 & 0.191  & 2.00 &  0.163 & 1.99 & 0.289 \\
			&20.557&-1.005 &-0.703 &2.00 &  0.091 & 1.97 & 0.853  & 2.00 & -0.111 & 2.00 & 0.446 \\
			&22.677&-0.977 &-0.210 &2.00 &  0.284 & 1.94 & 0.519  & 2.00 &  2.384 & 1.99 & 0.191 \\
			&25.222&-0.576 &-0.603 &1.98 & -0.586 & 1.93 & 1.632  & 2.00 &  0.214 & 1.99 & 0.464 \\
			&26.392&-0.823 &-1.664 &1.98 & -0.858 & 2.00 & 2.018  & 1.99 &  0.101 & 1.99 & 0.413 \\
			&26.942&-0.251 &-0.609 &1.99 &  0.070 & 1.92 & 2.136  & 2.00 & -0.046 & 1.96 & 0.550 \\
			&28.657&-0.261 &-0.645 &1.99 &  0.295 & 1.99 & 1.606  & 2.00 &  0.104 & 1.62 & 0.922 \\\hline
			$^5B_2$&23.347&-0.572 &-0.526 &5.98 &  0.292 & 4.65 & 1.800  & 5.99 &  0.145 & 5.99 & 0.479 \\
			\midrule
			\multirow{2}{*}{Low-lying states\tnote{f}}
			& MAE\tnote{g}    & 0.27  & 0.45& &0.16 & &0.27& &0.35 && 0.36 \\
			& MAX\tnote{h}    & 0.57  & 0.75& &0.59 & &0.68& &0.52 && 0.47 \\ \midrule
			\multirow{2}{*}{High-lying states\tnote{i}}
			& MAE\tnote{g}    & 0.62  & 0.80& &0.36 & &1.48& &0.57 && 0.65 \\
			& MAX\tnote{h}    & 1.16  & 1.71& &1.37 & &3.54& &2.93 && 2.74 \\ \midrule
			\multirow{2}{*}{All states\tnote{j}}
			& MAE\tnote{g}    & 0.52  & 0.70& &0.30 & &1.14& &0.51 && 0.57 \\
			& MAX\tnote{h}    & 1.16  & 1.71& &1.37 & &3.54& &2.93 && 2.74 \\
			\hline\hline
		\end{tabular}
		\begin{tablenotes}
			\item[a] Ground state canonical HF orbitals.
			\item[b] Projected $H_0^\alpha$ \eqref{ReferenceH0} with $f^\alpha$ \eqref{Fspin} in place of $F^\alpha$.
			\item[c] Semi-canonical SA-CASSCF orbitals.
			\item[d] $\mathcal{M}^\prime_0$ is selected from CAS(8,8) with $C_{\mathrm{min}}=10^{-2}$ (cf. Table \ref{table:ModelSpaceInfo}).
			\item[e] Expectation value of $S^2$.
			\item[f] 12 states in $[7, 18]$ eV.
			\item[g] Mean absolute error.
			\item[h] Maximum absolute error.
			\item[i] 31 states in $[18, 31]$ eV.
			\item[j] 43 states in $[7, 31]$ eV.
		\end{tablenotes}
	\end{threeparttable} \label{table:h2o_results}
\end{table}

\begin{table}
	\centering
	\caption{Mean absolute errors (MAE), maximum positive errors [MaxE(+)] and
minimal negative errors [MinE(-)]  (in eV)  of IN-GMS-SU-CCSD,
different variants of PASPT2 and CASSCF-SDSPT2 relative to iCIPT2 for the
vertical excitation energies of \ce{H2O}.
	}
	\begin{threeparttable}
		\begin{tabular}{lcccccccccccccccc}\hline\hline
			\multirow{1}{*}{Method}
			 & \multicolumn{3}{c}{12 states in $[7, 18]$ eV}
			 & \multicolumn{3}{c}{31 states in $[18, 31]$ eV} \\ \cline{2-7}
			                  & MAE  &MaxE(+)&MinE(-)
			                   & MAE & MaxE(+)& MinE(-)
			                  \\ \hline
			SU-CCSD             & 0.27 & 0.03 & -0.57 & 0.62 & 0.07 & -1.16 \\
			PASPT2\tnote{a,b}   & 0.51 & 0.17 & -0.74 & 1.18 & 0.42 & -2.10 \\
			PASPT2\tnote{a,c}   & 0.22 & 0.19 & -0.39 & 1.34 & 0.13 & -2.24 \\
			PASPT2\tnote{a,d}   & 0.45 & 0.20 & -0.75 & 0.80 & 0.53 & -1.71 \\
			PASPT2\tnote{a,f}   & 0.44 & 0.17 & -0.72 & 0.78 & 0.51 & -1.55 \\
            PASPT2\tnote{a,g}   & 0.28 & 0.18 & -0.46 & 0.27 & 0.39 & -0.65 \\
			PASPT2\tnote{h,d}   & 0.27 & 0.68 & -     & 1.48 & 3.54 & -     \\			
			PASPT2\tnote{a,d,i} & 0.16 & 0.59 & -     & 0.36 & 0.95 & -1.37 \\
			PASPT2\tnote{h,d,j} & 0.35 & 0.52 & -     & 0.57 & 2.93 & -0.92 \\
			SDSPT2              & 0.36 & 0.47 & -     & 0.65 & 2.74 & -1.45 \\
        \hline\hline
		\end{tabular}
		\begin{tablenotes}
			\item[a] Ground state canonical HF orbitals.
            \item[b] Diagonalization of $\bar{H}^{\mathrm{eff}[2]}$ \eqref{PASH2}.
            \item[c] Diagonalization of $\bar{H}^{\mathrm{eff}[2]}_Q$ \eqref{Q-open}.
            \item[d] Diagonalization of $\bar{H}^{\mathrm{eff}[2]}_X$ \eqref{HeffX}.
            \item[f] Diagonalization of $\bar{H}^{\mathrm{eff}[2]}_X$ \eqref{HeffX} under the LCUT approximation of the $\mathbf{t}$-amplitudes.
            \item[g] Diagonalization of $\bar{H}^{\mathrm{eff}[2]}_X$ \eqref{HeffX} with $M_C$ in place of $M_X$.
			\item[h] Semi-canonical SA-CASSCF orbitals.
			\item[i] Projected $H_0^\alpha$ \eqref{ReferenceH0} with $f^\alpha$ \eqref{Fspin} in place of $F^\alpha$.
			\item[j] $\mathcal{M}_0^\prime$ is selected from CAS(8,8) with $C_{\mathrm{min}}=10^{-2}$ (cf. Table \ref{table:ModelSpaceInfo}).
		\end{tablenotes}
	\end{threeparttable} \label{table:MAXMAE}
\end{table}

\subsection{Energy curves of \ce{N2}}
\begin{figure}[H]
	\centering
\begin{threeparttable}
		\begin{tabular}{cc}
			\includegraphics[width=0.5\linewidth]{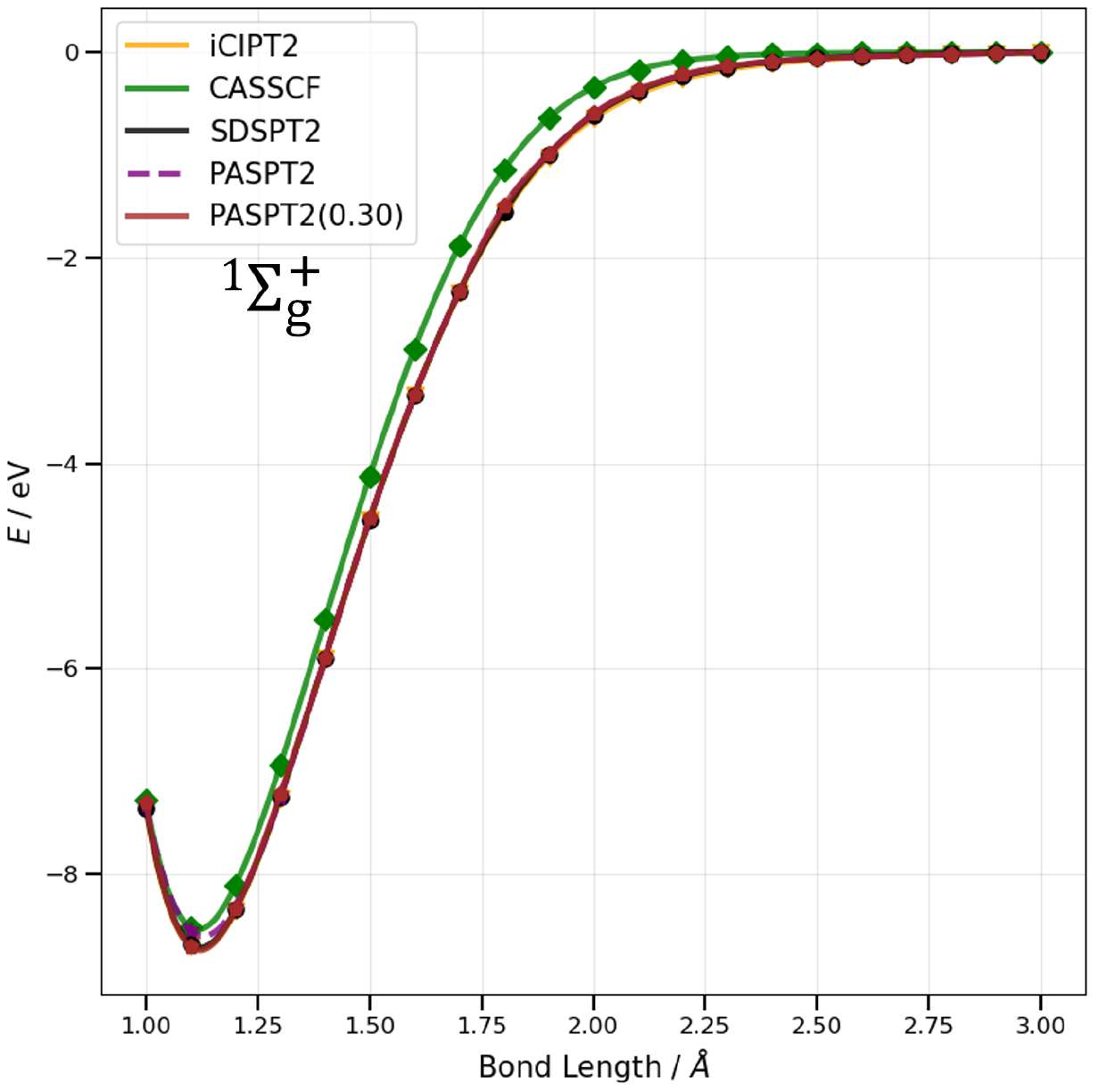} & \includegraphics[width=0.5\linewidth]{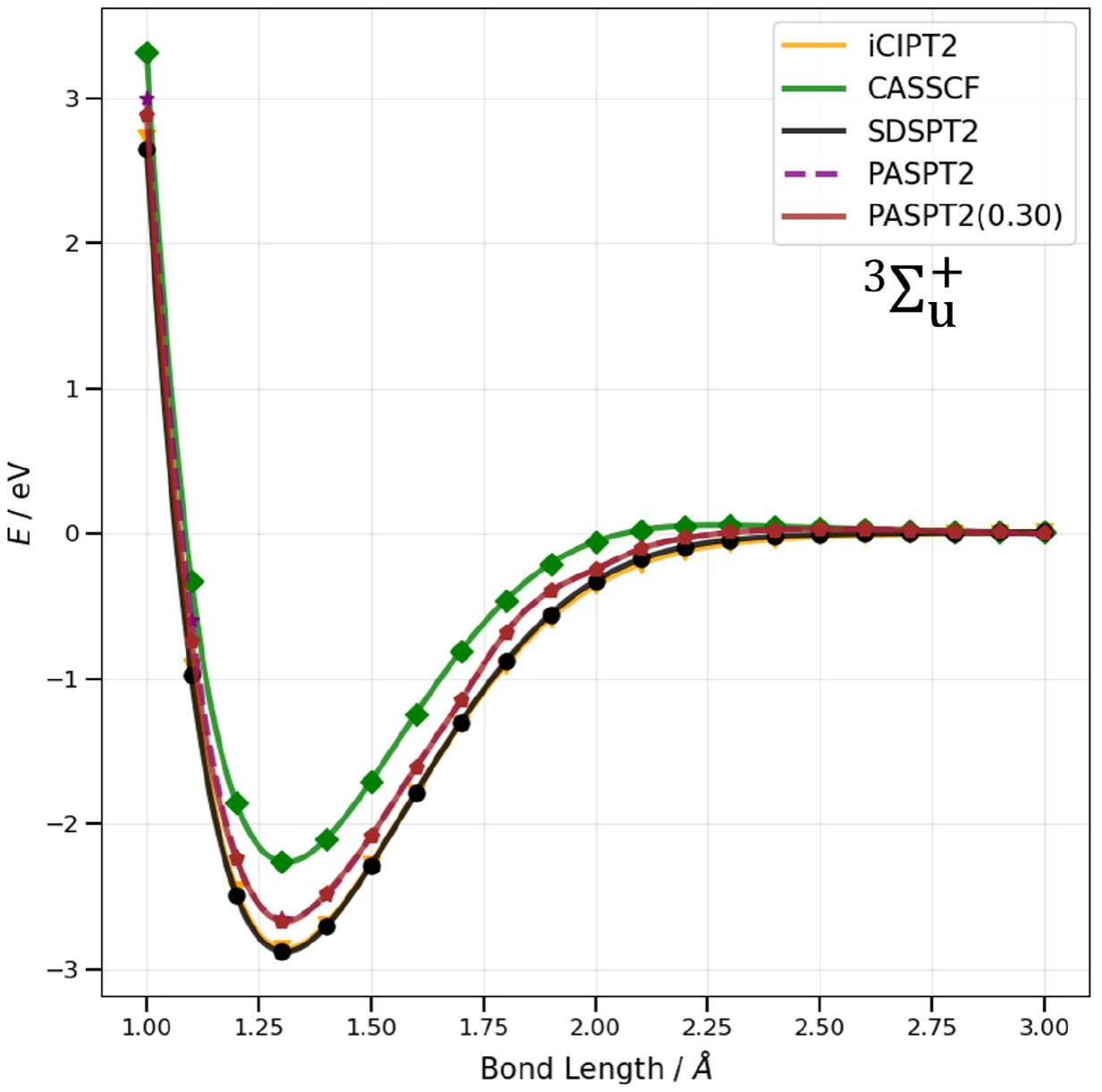} \\
			\includegraphics[width=0.5\linewidth]{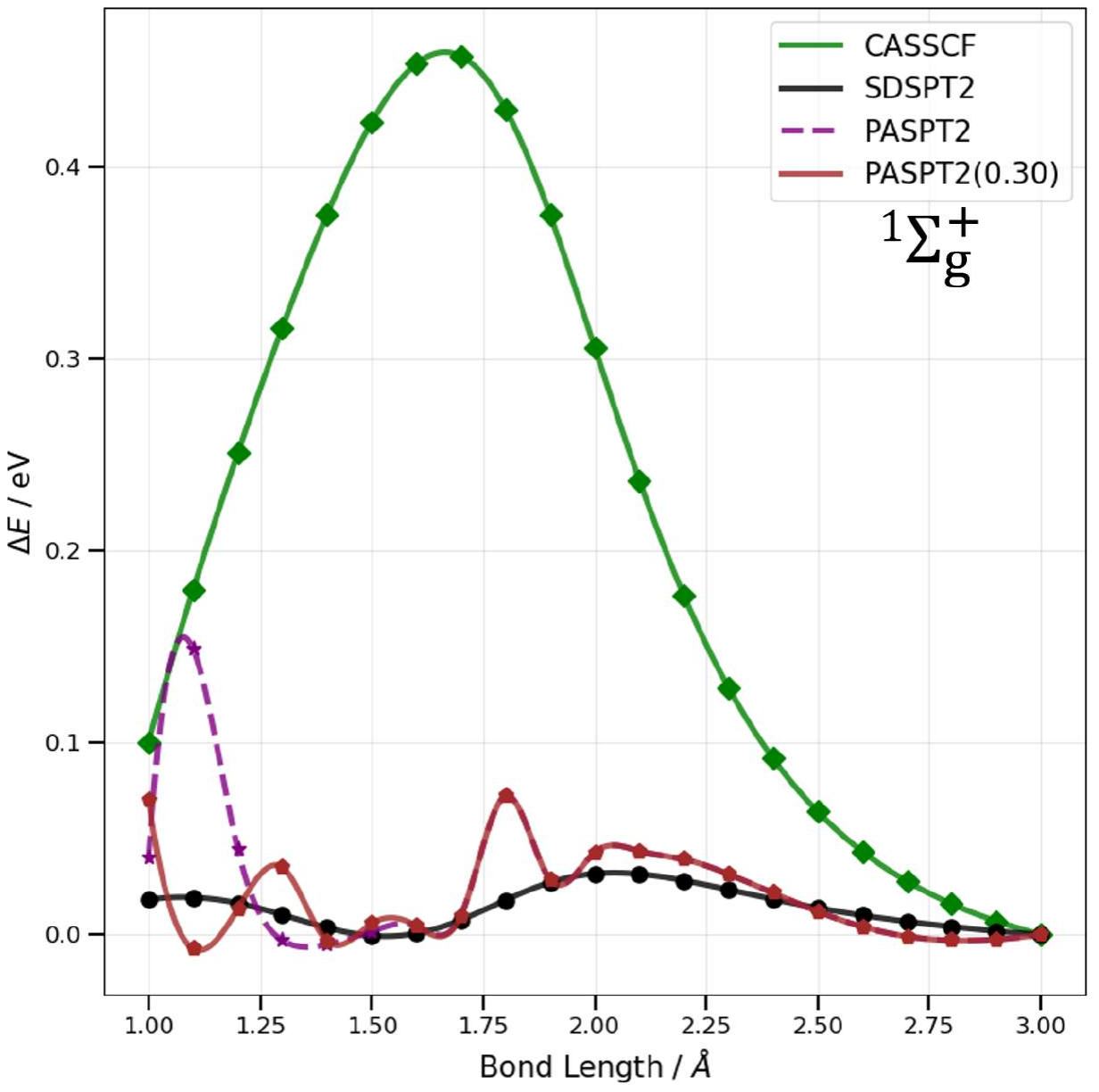} & \includegraphics[width=0.5\linewidth]{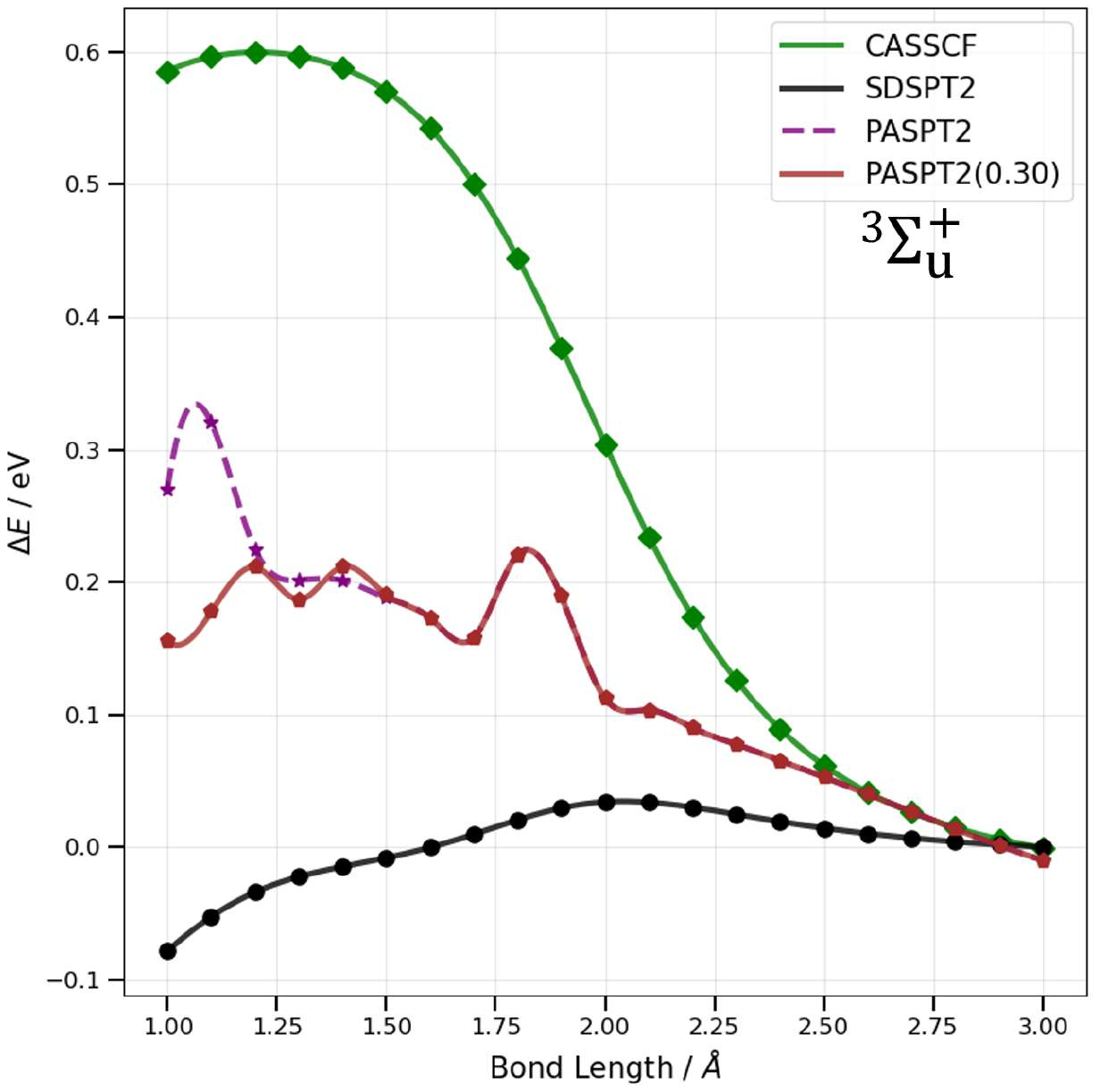}
		\end{tabular}
\end{threeparttable}
	\caption{Potential energy curves of the $^1\Sigma_g^+$ and $^3\Sigma_u^+$ states of \ce{N2}.
 		First row: full PECs by CASSCF, PASPT2, PASPT2(0.3), CASSCF-SDSPT2 and iCIPT2 with the cc-pVDZ basis;
		second row: deviations of CASSCF, PASPT2, PASPT2(0.3) and CASSCF-SDSPT2 from iCIPT2.
The iCIPT2 energy at 3.0~\AA~is taken as the zero-energy point for all curves.
`PASPT2(0.3)' means the exclusion of external excitations with uncoupled amplitudes larger than 0.3.
		 }
	\label{fig:n2_curve}
\end{figure}

\begin{figure}[H]
	\centering
\begin{threeparttable}
		\begin{tabular}{cc}
			\includegraphics[width=0.5\linewidth]{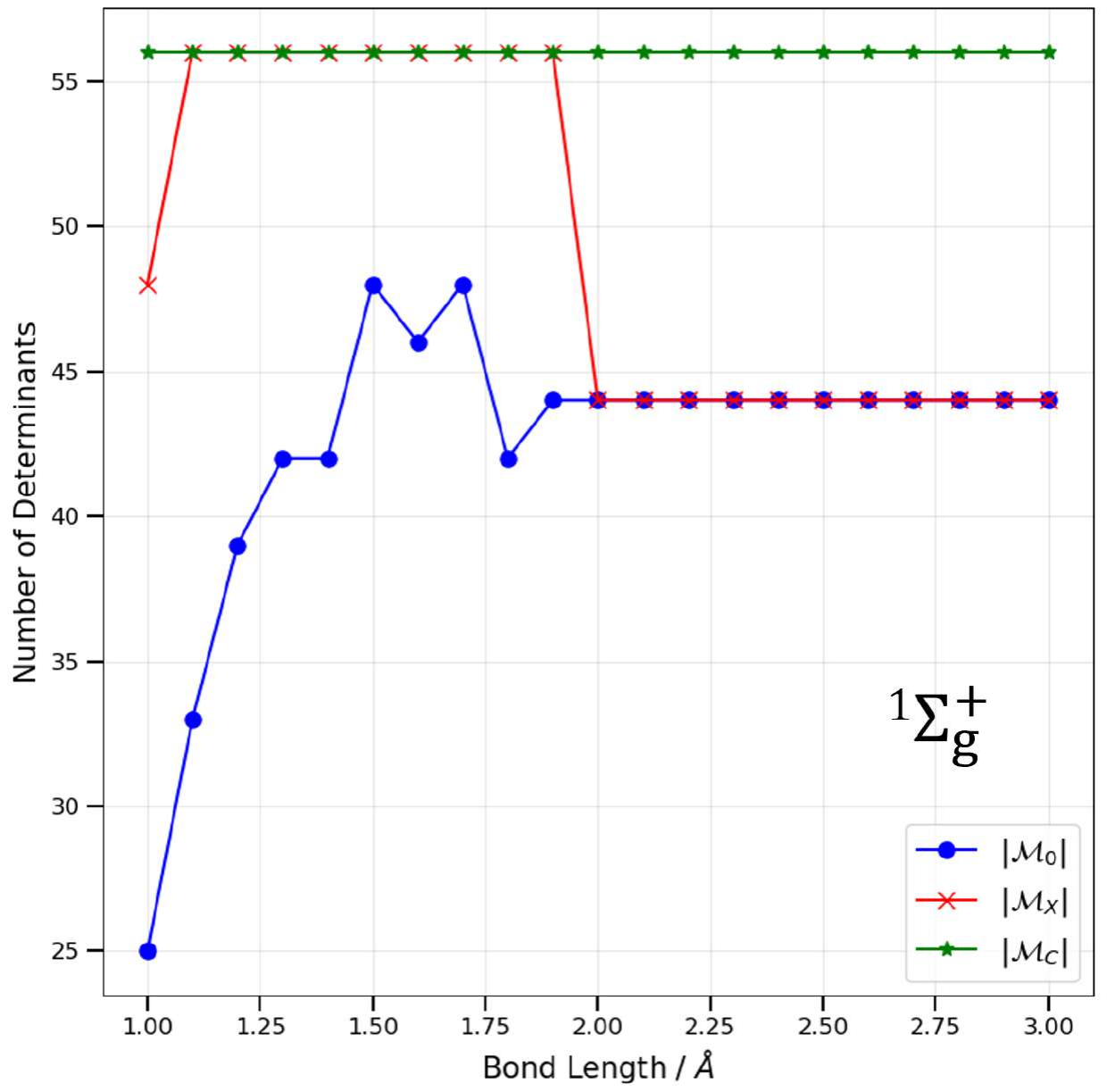} & \includegraphics[width=0.5\linewidth]{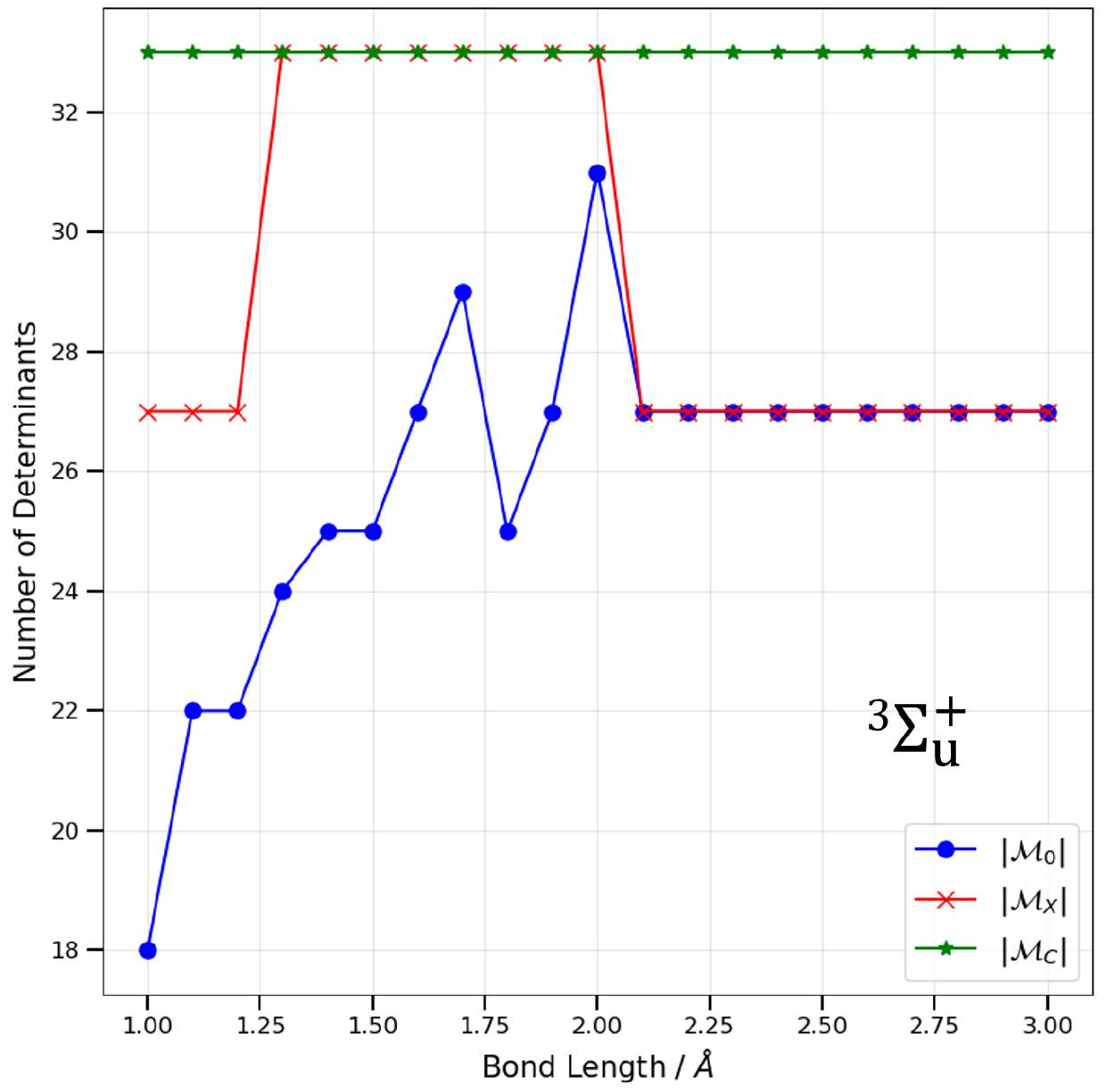}
		\end{tabular}
\end{threeparttable}
	\caption{Variation of $|\mathcal{M}_0|$, $|\mathcal{M}_X|$ and $|\mathcal{M}_C|$ along the interatomic distance of \ce{N2}.
		 }
	\label{fig:n2_curveM0}
\end{figure}

The potential energy curves (PEC) of the ground ($^1\Sigma_g^+$) and first excited ($^3\Sigma_u^+$) states of
the \ce{N2} molecule are a classic test ground for multi-reference methods.
It is especially challenging for methods (e.g., PASPT2) that do not work with a CAS.
The minimal CAS for the correct dissociation of \ce{N2} is CAS(6,6) that involves
one $\sigma$ bond and two $\pi$ bonds as well as their anti-bonding counterparts.
After state-specific CASSCF(6,6) calculations, only those NEDs with coefficients larger than $10^{-2}$
in absolute value were retained in the model space $\mathcal{M}_0$, which is then
extended automatically to form a closed model space $\mathcal{M}_X$ for PASPT2.
It is first seen from Table \ref{NeReDZ} that the PASPT2 molecular spectroscopic constants
(equilibrium bond length $R_e$, harmonic vibrational frequency $\omega_e$, dissociation energy $D_e$)
are in fairly good agreement with those by iCIPT2 (NB: as can be seen from Table S2 in the Supporting
Information, iCIPT2 deviates from full configuration interaction\cite{N2FCI} by less than 0.01 m$E_h$ for the energies
across all interatomic distances of \ce{N2}). However, a close inspection of the PECs (cf. the second row of Fig. \ref{fig:n2_curve})
reveals that the PASPT2 PECs in the bonding regions (1.0$\sim$1.3~\AA) of both $^1\Sigma_g^+$ and $^3\Sigma_u^+$
are not very accurate. This stems from the fact that some uncoupled amplitudes $t^0_{l\alpha}$ are
unduely large in this region (which does not appear in the case of \ce{H2O}). They are associated to some single excitations
from the reference NEDs (which have the smallest coefficients
in the wave functions obtained by diagonalizing $PHP$) to $\mathcal{M}_C^\perp$,
instead of to $\bar{\mathcal{M}}_X$ ($=\mathcal{M}_C\backslash\mathcal{M}_X$).
Since the corresponding energy denominators are not small at all (>0.1 $E_h$),
such improper amplitudes must be ascribed to the violation of the Brillouin conditions.
As said before, such external excitations can either be reduced
by introducing a suitable shift to the energy denominators $\Delta E_{l\alpha}$ or simply dumped (which
amounts to using an infinitely large shift). After some experimentations,
the latter is adopted here, that is, those external functions with $t^0_{l\alpha}$ larger than 0.3
are simply removed. This amounts to eliminating those components of the first-order correction to a reference NED
with weights larger than ca. 10\%. As characterized by the much reduced non-parallelity error,
the PECs in the bonding regions are indeed improved this way, which is accompanied by
improved molecular spectroscopic constants as shown in Table \ref{NeReDZ}.
The efficacy of PASPT2 for smooth PECs is gratifying particularly when observing that
both $|\mathcal{M}_0|$ and $|\mathcal{M}_X|$ vary along the interatomic distance
and $\mathcal{M}_X$ may be identical to $\mathcal{M}_C$ or $\mathcal{M}_0$ itself, see Fig.
\ref{fig:n2_curveM0}.

To see the performance of PASPT2 for a larger basis set, the calculations of the PECs of \ce{N2} were repeated with the cc-pVTZ basis\cite{Dunning1989}.
To establish the reference data, the iCIPT2 calculations were performed with 5 values for $C_{\text{min}}$ (i.e., $\{15,9,7,5,2\}\times10^{-6}$)
and then extrapolated by linear fits of the $E_{tot}$ vs $|E_c^{(2)}|$ plots.
Here, $C_{\text{min}}$ is the threshold for the selection of configurations, $E_{tot}$ is the total energy,
whereas $E_c^{(2)}$ is the second-order perturbation correction (for more details, see Ref. \citenum{iCIPT2New}).
The so-extrapolated point-wise total energies for the ground and first excited states of \ce{N2}, along with
those by PASPT2 and CASSCF-SDSPT2, are documented in Tables S3 and S4,
respectively, and further plotted in Fig. S2 in the Supporting Information. Here,
it should be mentioned that the existence of external excitations of large uncoupled amplitudes
renders the iterations of the PASPT2 amplitude equations difficult to converge
and hence has to be removed. Nevertheless,
it can be seen from Table \ref{NeReTZ} that the PASPT2 molecular spectroscopic constants remains in good agreement with those by iCIPT2.

Finally, it deserves to be pointed out that CASSCF-SDSPT2 performs extremely well
for the PECs of both the $^1\Sigma_g^+$ and $^3\Sigma_u^+$ states of \ce{N2}.

\begin{table}
	\small
	\centering
	\caption{Deviations of CASSCF, PASPT2 and CASSCF-SDSPT2 from iCIPT2 for the molecular
spectroscopic constants (equilibrium bond length $R_e$, harmonic vibrational frequency $\omega_e$, dissociation energy $D_e$)
for the ground ($^1\Sigma_g^+$) and first excited ($^3\Sigma_u^+$) states of \ce{N2} with the cc-pVDZ basis.
	 }
	\begin{threeparttable}
		\begin{tabular}{ccrrrrrr}\hline\hline
	State & Method & $R_e$/\AA & $\omega_e$/cm$^{-1}$ & $D_e$ / eV & NPE\tnote{a} / meV  & NPE\tnote{b} / meV  & NPE\tnote{c} / meV  \\\toprule
$^1\Sigma_g^+$	&iCIPT2\tnote{d}&1.125&1954.0&8.76&-&-&-\\
	&CASSCF&-0.004&26.2 &-0.20&392&231&8 \\
	&PASPT2& 0.005&-22.8&-0.13&59&28&7 \\
	&PASPT2\tnote{e}&0.000&15.7&0.01&31&28&7 \\
	&SDSPT2&0.000&-0.2&-0.02&4&21&2 \\ \hline
$^3\Sigma_u^+$
    &iCIPT2&1.312&1559.9&2.86&-&	-&	-\\
	&CASSCF&0.001&-18.0&-0.60&574&235&8 \\
	&PASPT2&-0.001&8.8&-0.20&254&151&33 \\
	&PASPT2\tnote{e}&-0.002&0.8&-0.19&185&153&33 \\
	&SDSPT2           &-0.002&3.7&0.02&11&23&2 \\ \hline\hline
	\end{tabular}
	\begin{tablenotes}
	\item[a] Non-parallelity error for bond lengths 1.0 to 1.3 \AA.
	\item[b] Non-parallelity error for bond lengths 1.4 to 2.3 \AA.
	\item[c] Non-parallelity error for bond lengths 2.4 to 3.0 \AA. For the ground state, the distance 3.0~\AA~ is excluded.
	\item[d] $C_{\text{min}}=10^{-6}$.
    \item[e] External excitations with uncoupled amplitudes $t^0_{l\alpha}$ larger than 0.3 in absolute value are removed.
	\end{tablenotes}
	\end{threeparttable}\label{NeReDZ}
\end{table}

\begin{table}
	\small
	\centering
	\caption{Deviations of CASSCF, PASPT2 and CASSCF-SDSPT2 from iCIPT2 for the molecular
		spectroscopic constants (equilibrium bond length $R_e$, harmonic vibrational frequency $\omega_e$, dissociation energy $D_e$)
		for the ground ($^1\Sigma_g^+$) and first excited ($^3\Sigma_u^+$) states of \ce{N2} with the cc-pVTZ basis.
	}
	\begin{threeparttable}
		\begin{tabular}{ccrrrrrr}\hline\hline
			State & Method & $R_e$/\AA & $\omega_e$/cm$^{-1}$ & $D_e$ / eV & NPE\tnote{a} / meV  & NPE\tnote{b} / meV  & NPE\tnote{c} / meV  \\\toprule
			$^1\Sigma_g^+$	&iCIPT2\tnote{d} &  1.112 &1896.6&9.53&-&-&-\\
			                &CASSCF &  0.001 &38.4 &-0.68&680&497&55 \\
			                &PASPT2\tnote{e}& -0.002 &-16.3&-0.02&35&11&14 \\
			                &SDSPT2         &  0.001 &  0.4&-0.10&95&49&12 \\ \hline
			$^3\Sigma_u^+$
			                &iCIPT2\tnote{d} & 1.295 &1562.6&3.39&-&	-&	-\\
			                &CASSCF & 0.012 &-38.3&-0.94&1068&541&53 \\
			                &PASPT2\tnote{e}& -0.003 &-10.1&-0.27&233&174&35 \\
			                &SDSPT2         & -0.001 &2.3  &-0.03&17&46&12 \\ \hline\hline
		\end{tabular}
		\begin{tablenotes}
			\item[a] Non-parallelity error for bond lengths 1.0 to 1.3 \AA.
			\item[b] Non-parallelity error for bond lengths 1.4 to 2.3 \AA.
			\item[c] Non-parallelity error for bond lengths 2.4 to 3.0 \AA. For the ground state, the distance 3.0~\AA~ is not included.
			\item[d] Linearly extrapolated results (cf. Tables S2 and S3).
			\item[e] External excitations with uncoupled amplitudes $t^0_{l\alpha}$ larger than 0.3 in absolute value are removed.
		\end{tablenotes}
	\end{threeparttable}\label{NeReTZ}
\end{table}

\section{Conclusions and outlook}\label{Conclusion}
A novel MS-MRPT2, PASPT2, has been proposed, implemented and tested against prototypical systems.
It is based on the Jeziorski-Monkhorst wave operator as well as
the Li-Paldus connectivity condition associated with the intermediate normalization for a partial/incomplete active space.
Although taken as a start for formulating PASPT2, the IN-GMS-SU-CC theory of Li and Paldus is found to be \emph{not} size-extensive, contrary to the original belief.
By contrast, the termwise connectedness of the PASPT2 amplitudes has been achieved by introducing a reference-specific zeroth-order Hamiltonian
consisting of a class-diagonal inactive term and a nonhermitian active term. Moreover, the
effective/intermediate Hamiltonian has also been made connected and closed, so that PASPT2 is strictly size-extensive.
It is also size-consistent when the PAS is chosen to be the direct product of those of non-interacting subsystems.
To the best of our knowledge, PASPT2 is up to date the only size-extensive, size-consistent and intruder-free PAS-based MS-MRPT2 under the intermediate normalization.
The next step is to make PASPT2 spin adapted, which will be reported elsewhere.

\section*{Acknowledgments}
This work was supported by the National Natural Science Foundation of China (Grant Nos. 22373057 and 22503051).

\section*{Supporting Information}
Comparison of different model spaces for \ce{H2O}; point-wise energies of \ce{N2} by various methods and basis sets.

\section*{Conflicts of interest}
There are no conflicts to declare.

\bibliography{iCI}

\end{document}